\documentclass[final,5p,times,twocolumn]{elsarticle}

\makeatletter
\def\ps@pprintTitle{%
 \let\@oddhead\@empty
 \let\@evenhead\@empty
 \def\@oddfoot{\centerline{\hspace*{0.5cm}Accepted for publication in JHEAP}}%
 \let\@evenfoot\@oddfoot
}
\makeatother

\usepackage{amssymb}
\usepackage{lipsum}
\usepackage{amsmath} 
\usepackage[normalem]{ulem}
\usepackage{amsthm}

\usepackage{xcolor}

\usepackage{subcaption} 

\usepackage{hyperref}
\usepackage{doi}
\usepackage{bm}
\usepackage{mathtools}
\usepackage{graphicx}
\usepackage{natbib}

\usepackage[switch]{lineno}

\journal{High Energy Astrophysics}

\begin{document}

\begin{frontmatter}



\title{TeV gamma-ray spectral spikes produced by magnetic reconnection in blazar jets:\\ 
the case of the 2014 high state of Markarian 501}


\author[first]{J. G. Giesbrecht Formiga Paiva}
\affiliation[first]{organization={Centro Brasileiro de Pesquisas F\'isicas},
            addressline={Xavier Sigaud st. 150}, 
            city={Rio de Janeiro},
            postcode={22290-180}, 
            state={RJ},
            country={Brazil}}
            
\author[second]{E. M. de Gouveia Dal Pino}
\affiliation[second]{organization={Universidade de S\~{a}o Paulo, Instituto de Astronomia, Geof\'{i}sica e Ci\^{e}ncias Atmosf\'{e}ricas},
            addressline={1226 Matão Street}, 
            city={S\~{a}o Paulo},
            postcode={05508-090}, 
            state={SP},
            country={Brazil}}

\author[first]{J. C. Rodr\'iguez-Ram\'irez}  

\author[first]{U. Barres de Almeida}

\author[third]{G. B. D\'iaz-Cort\'es}          \affiliation[third]{organization={Instituto Mexicano del Petr\'oleo (IMP)},
            addressline={Eje Central L\'azaro C\'ardenas 152, Col. San Bartolo Atepehuacan}, 
            city={Mexico City},
            postcode={07730}, 
            country={Mexico}}

\begin{abstract}
Multi-wavelength monitoring of the flaring blazar Markarian 501 during July 2014
revealed a TeV gamma-ray spike feature with 3-4$\sigma$ significance and coincident with a prominent enhancement in its  X-ray flux.
The appearance of this spectral feature strongly suggests the presence of an extra emission component in addition to the usual one-zone SSC scenario.
Several possible explanations for the origin of this novel behavior have been discussed, including stochastic particle acceleration, magnetospheric vacuum gap, and pion decay.
In this paper, we show that the TeV narrow feature, simultaneous with an increase in the X-ray flux, can be produced with two leptonic emission regions in a jet undergoing magnetic reconnection energy dissipation along its propagation axis.
In this scenario, the stable spectral components are produced in the region of maximum magnetic dissipation.
A second region produces a flare upstream in the jet 
in a slower, more magnetized, and much smaller region compared to the stable one, which is responsible for increasing the X-ray flux and producing the TeV spike.
The macroscopic properties of these two emission regions are consistent with a magnetically striped jet model discussed in previous works, where the acceleration of the jet flow and its non-thermal emission is driven by turbulent-induced magnetic reconnection.
We employ this jet-reconnection scenario to model the 2014 high state of the blazar Markarian 501, considering the sequence of SED datasets corresponding to 
MJD  56855.91, 56857.98, 56858.98, and  56859.97, with the second dataset being the one that exhibits 
the TeV spike.
\end{abstract}




\end{frontmatter}




\section{Introduction}
\label{introduction}

Blazars are extragalactic sources that exhibit powerful broadband emission spanning from radio wavelengths to $\gamma$ rays, generally interpreted as arising from a collimated, relativistic plasma outflow (a jet) oriented close to our line of sight.
The broadband spectral energy distribution (SED) of blazars is characteristically double-humped and dominated by non-thermal emission produced by relativistic charged particles. Sources that lack strong optical emission lines are classified as BL Lacertae objects (BL Lacs), whereas those that exhibit prominent broad emission lines and a clear thermal contribution from the accretion disc are referred to as flat-spectrum radio quasars (FSRQs).
Blazars are also commonly classified according to the frequency of the synchrotron peak, $\nu_{\rm s}$, in their SED: low-synchrotron-peaked objects (LSPs, $\nu_{\rm s} < 10^{14},\text{Hz}$), intermediate-synchrotron-peaked objects (ISPs, $10^{14},\text{Hz} < \nu_{\rm s} < 10^{16},\text{Hz}$), and high-synchrotron-peaked objects (HSPs, $\nu_{\rm s} > 10^{16},\text{Hz}$) \cite{abdo2010spectral}. In recent years, with the expansion of blazar catalogues, a further subclass of extreme high-synchrotron-peaked objects (EHSPs), characterised by $\nu_{\rm s} \gtrsim 10^{17},\text{Hz}$, has been recognised \cite{biteau2020progress,costamante2001extreme}.

First detected in the very-high-energy (VHE; E$>100$ GeV) range by Whipple Observatory in 1995, Markarian 501 (Mrk 501) is a blazar classified as BL Lac type. 
With a redshift of 0.034, it is also considered one of the closest objects in the extragalactic sky. 
Regarding the peak of its low energy spectral bump, Mrk\,501, is generally classified as an HSP but has occasionally entered EHSP states during periods of intense VHE activity, with its synchrotron peak shifting toward or beyond $\sim10^{17}\,$Hz \cite{abdo2010spectral,costamante2001extreme}. 
These spectral transitions, the absence of strong thermal components, its peculiar high-energy emission features,
as well as its relative proximity make Mrk\,501 a key target for probing high-energy processes blazar jets.

Theoretically, blazar jets are understood as 
energised by a spinning, accreting supermassive black hole (SMBH) at the centre of the host galaxy. 
The Blandford–Znajek (BZ) process \textbf{\cite{blandford1977electromagnetic}} is widely regarded as a dominant mechanism for launching blazar jets. In this picture, the outflow is born (magnetically) Poynting-dominated and must be accelerated to large bulk Lorentz factors as it propagates away from the central engine. The detailed jet structure and the microphysics of particle acceleration, however, remain not completely understood and are the subject of active research.
The location in the blazar jet where the VHE emission is produced, is likewise uncertain. A transition from magnetically dominated to kinetically dominated flow is expected, implying substantial magnetic-energy dissipation—a natural driver of particle acceleration and radiation. Fast magnetic reconnection provides such a dissipation route and has been extensively investigated with both MHD and PIC simulations (e.g., 
\cite{kowal2012particle,del2016properties,singh2016spatial,kadowaki2018mhd,guo2019determining,sironi2014relativistic,comisso2018particle,comisso2019interplay,beresnyak2016first,ripperda2017reconnection,zhang2023particle,medina2021particle,medina2023particle}). In MHD flows, turbulence is a powerful  driver of fast reconnection 
\cite{lazarian1999reconnection, eyink2013flux}, and particles energized within reconnection layers (current sheets) undergo a first-order Fermi process 
\cite{gouveia2005production}.
 The efficiency of turbulent-driven reconnection acceleration in producing very and ultra-high energy cosmic rays (UHECRs)  has been successfully tested, particularly in relativistic AGN-jet simulations %
 \cite{medina2021particle,pino2024particle}.
In this work, we apply this framework to interpret the broadband emission of Mrk~501—and, in particular, 
the enhancement of its X-ray/VHE emissions
following the formalism of 
\cite{dgdp_2025}.

The multi-wavelength campaign on Mrk\,501 in 2014 revealed its most active VHE period since 1997, as well as the most extreme X-ray activity ever detected from the source with \textit{Swift}-XRT \cite{MAGIC_2020}. Notably, this campaign captured a transient, unusual behaviour: a narrow spectral feature at $\sim 3$~TeV, temporally coincident with an enhancement of the X-ray flux, suggesting the presence of an additional emission component superimposed on the regular source emission.
While a number of possible interpretations have been proposed, including hadronic and multi-zone emission scenarios (discussed in detail in the next section), the origin of the TeV feature remains debated.

In this work, we propose that the TeV spike can be interpreted as arising from two distinct emission regions powered by magnetic reconnection in a jet transitioning from a magnetically dominated to a kinetically dominated flow.
Multiple and simultaneous emission zones in astrophysical jets are not unexpected, as observed in both Galactic and extragalactic jets \cite{Noriega_2020, Fender_1999, Stirling_2002, Harris_2006}. For instance, it is well established that radio emission at GHz frequencies from blazars sources originates in more extended regions of the jet, downstream of the compact blazar zone \cite{Jorstad_2005}. Blazars also exhibit variability across different time scales and wavelengths \cite{Prince_2019}, further indicating that multiple emitting zones contribute to their spectra.

The two-zone blazar model proposed here consists of a quasi-steady base component and a transient component, the latter arising from a smaller, more magnetized region located upstream along the jet (see Fig.\ref{fig:sketch}). We place the stable component where the jet undergoes its peak magnetic dissipation during the magnetic-to-kinetic energy conversion, and we parameterize the jet and radiation zones with the striped-jet framework of \cite{Giannios_2019}, in which magnetic reconnection drives this transition. In our model, reconnection is fast due to turbulence embedded in the flow, and the setup extends the scheme of de Gouveia Dal Pino et al.\cite{dgdp_2025}. We restrict the radiative treatment to leptonic emission in both emission zones and neglect external photon fields, a scenario consistent with typical BL~Lac spectra.

This paper is organized as follows. In Section~2, we describe the datasets used, while also showing the hypotheses given so far for the SED's spike of Mrk 501 in 2014. Section~3 presents the analytic framework of the proposed emission scenario. In Section~4, we apply the model to four SED datasets of Mrk~501—sampling epochs before, during, and after the emergence of the narrow TeV feature. Section~5 summarizes the results and discusses the implications of the model.

\section{The historical background of Mrk 501's SED feature in July 2014}

In July 2014, the MAGIC Collaboration conducted a multi-wavelength observational campaign of Mrk 501 \cite{MAGIC_2020}, covering energies from radio to VHE gamma rays.
On July 19, 2014, the source exhibited a pronounced spectral spike at approximately $\sim 3\,\mathrm{TeV}$ \cite{MAGIC_2020}, coinciding with the highest X-ray activity detected by \textit{Swift}-XRT on the same day \cite{MAGIC_2020}.
In this context, the VHE gamma-ray SED was best described by either a broad log-parabola combined with a narrow Gaussian-like component or an exponential log-parabolic shape, rather than a simple power-law or single log-parabola \cite{MAGIC_2020}. Statistically, the spectral spike was detected with a significance of 3–4 $\sigma$ \cite{MAGIC_2020}.

To date, the models proposed to explain the origin of this spectral spike include: (i) a pile-up in the electron energy distribution (EED) \cite{MAGIC_2020,hu2021narrow}, caused by stochastic or shock acceleration, (ii) emission from an additional synchrotron self-Compton (SSC) zone \cite{MAGIC_2020}, (iii) contributions from a magnetospheric vacuum gap \cite{MAGIC_2020,wendel2021electron}, and (iv) pion decay \cite{petropoulou2024tev}. While all these scenarios can reproduce the observed feature, each requires specific assumptions and constraints. Notably, these scenarios focus primarily on reproducing the conditions on the day the spike was observed.

According to \cite{MAGIC_2020}, the most consistent explanation for the EED pile-up would involve two successive injections of mono-energetic electrons within the same region, separated by a time delay. In this scenario, the first population undergoes thermalization, forming a relativistic Maxwellian distribution around the equilibrium energy—defined as the energy where the acceleration timescale equals the cooling timescale \cite{MAGIC_2020}. This process naturally produces a spectral peak in the SED at approximately 3 TeV. Similarly, \cite{hu2021narrow} demonstrated that the spectrum can also be reproduced when shock acceleration dominates over stochastic acceleration, based on solutions of the Fokker–Planck equation for the electron distribution.
As mentioned earlier, this explanation—whether via thermalization or stochastic acceleration— considered only
the SED dataset corresponding to the day of the spike. Within this framework, it is challenging to track the evolution of the EED pile-up over several days, particularly given that \cite{MAGIC_2020} reports hints of the TeV narrow feature in the spectra both before and after the event, albeit less prominently than on July 19th.

The second proposed explanation involves introducing an additional SSC emission zone within a larger X-ray to gamma-ray emission region, with the smaller zone being responsible for the observed spectral spike \cite{MAGIC_2020}. However, this approach significantly increases the number of model parameters and requires adopting substantially different Doppler factors for the two zones, regardless of whether they are co-spatial or not \cite{MAGIC_2020}.

Considering the inverse-Compton (IC) cascades forming within the magnetospheric vacuum gap, it has been proposed that low-energy electrons from the accretion disk and/or interstellar gas clouds may enter the gap, triggering the cascade \cite{MAGIC_2020, wendel2021electron}. For this scenario to be viable, the accretion disk or surrounding gas must produce a low flux of photoemission lines \cite{MAGIC_2020, wendel2021electron}. However, other studies suggest that the jet emission in BL Lac objects can dominate over the recombination-line emission, effectively obscuring such features from detection \cite{MAGIC_2020}.

Finally, \cite{petropoulou2024tev} proposes a leptohadronic model involving combined SSC and $\pi^0$ decay emissions, where neutral pions are produced through interactions between low-energy protons and high-energy photons in the hard X-ray to soft gamma-ray range. This process naturally yields a narrow spectral peak around 3 TeV. The model specifically focuses on the spectral evolution during both pre-flare and flaring phases, but only considering the SED dataset of the day when the TeV feature is observed. 
The cooling time scale of the particles is not explicitly demonstrated in this study, 
which further complicates the description of the model’s day-to-day evolution.

As discussed earlier, the scenario proposed in this paper aims to explain the origin of the narrow TeV spike with a simultaneous enhancement of the X-ray flux through two emission zones driven by magnetic reconnection using a minimal-parameter model, while also
ensuring that the cooling timescales of the emitting particles remain consistent with the observed flux variability of the source.
To this end, we model both emission zones within a physically motivated jet structure that transitions from a magnetically dominated to a kinetically dominated flow \cite{Giannios_2019}, rather than arbitrarily parameterizing the properties of the emitting regions, and we consider SED datasets obtained in the days before and after the detection of the peculiar TeV spike.
The microphysical treatment, on the other hand,
is motivated by recent 3D MHD simulations of relativistic jets including test particles, which demonstrate that turbulence-driven magnetic reconnection can efficiently accelerate particles to ultra-high energies \cite{medina2023particle,pino2024particle}, as well as by the successful application of this framework to reproduce the neutrino and gamma-ray emission of the blazar TXS~0506+256 \citep{dgdp_2025}.

\section{The jet emission scenario}
\label{sec:basic_model}

In the scenario proposed here, the blazar emission originates from two distinct locations in the jet, situated near the transition from magnetic to a kinetic dominated flow. In this region, instabilities are expected to trigger turbulence-driven magnetic reconnection, leading to efficient particle acceleration, as demonstrated by 3D MHD simulations including test particles \cite{medina2021particle,medina2023particle,pino2024particle}. To model this process, we adopt an analytical approach based on the striped-jet reconnection model proposed by \cite{Giannios_2019}, further extended in \cite{dgdp_2025} to account for the effects of turbulence on particle acceleration.

\begin{figure}
   \centering
   \includegraphics[width=\hsize]{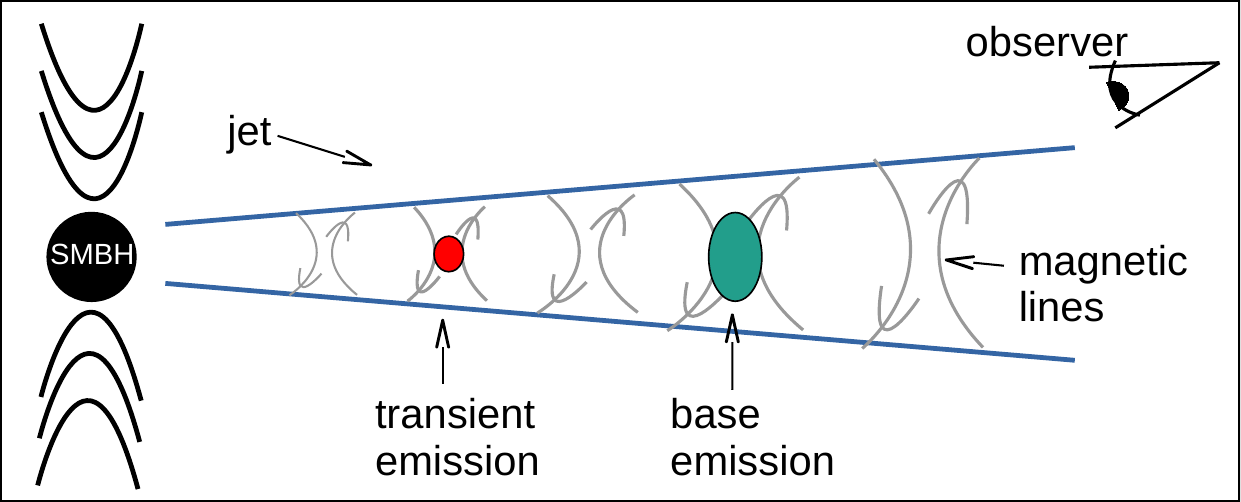}
\caption{Schematic illustration of the emission scenario considered in this work. Dimensions are not to scale.}
         \label{fig:sketch}
\end{figure}

In turbulence-driven magnetic reconnection, particles are primarily accelerated within reconnection layers through a first-order Fermi process, in which the relative energy gain per interaction scales as $\Delta E / E \propto v_{\mathrm{rec}} / c$ \citep{dgdp_2005}. Consequently, an efficient acceleration mechanism requires a fast reconnection rate ($v_{\mathrm{rec}}$). Such fast reconnection can be triggered by turbulence in MHD flows \citep{lazarian1999reconnection}, and recent studies have demonstrated its dominance over alternative mechanisms, such as the tearing mode instability \cite{vicentin2025investigating, morillo2025magnetic}.
Figure~\ref{fig:sketch} presents a schematic view of the magnetic reconnection stripes, where we assume particles are accelerated via the first-order Fermi process within these reconnection layers with embedded turbulence, as in (e.g., Kowal et al. (2012) \cite{kowal2012particle}, Medina-Torrejon et al. 2021 \cite{medina2021particle}; 2023 \cite{medina2023particle}; de Gouveia Dal Pino and Medina-Torrejon (2024) \cite{pino2024particle}).

At macroscopic scales, according to the framework of \cite{Giannios_2019},  the conversion of magnetic to kinetic energy in the jet is driven by magnetic reconnection between poloidal stripes of opposite polarity, oriented perpendicular to the jet axis. Such a striped magnetic structure may naturally arise from the rotation of an SMBH accretion disc threaded by magnetic loops.
\cite{Giannios_2019} assume that the stripe widths, $l$, follow a power-law distribution $\propto (l/l_\mathrm{min})^{-a}$, where $a$ is the spectral index and $l_\mathrm{min}$ the minimum stripe width.

The gradual dissipation of magnetic energy causes the magnetic field strength to decrease with distance $s$ from the SMBH, while the bulk Lorentz factor of the flow increases up to an asymptotic value $\Gamma_\infty$.
This transition is driven by the magnetic dissipation power, $P_\mathrm{diss}$, which rises with $s$ and reaches a maximum at
\begin{equation}
s_\mathrm{max} = \frac{l_\mathrm{min}\Gamma_\infty^2}{6\xi_\mathrm{rec}},
\label{smax}
\end{equation}
after which it gradually declines. In the above expression, $\xi_\mathrm{rec} = v_\mathrm{rec}/v_\mathrm{A}$ is the reconnection rate, which we link to the particle acceleration timescale described in Section~\ref{subsec:atime}.

The jet structure as a function of distance $s$ from the central engine is determined by specifying the parameters
$\Gamma_\infty$,
$L_\mathrm{j}$,
$\xi_\mathrm{rec}$,
$l_\mathrm{min}$, and
$a$.
Examples of the resulting profiles for the co-moving magnetic field $B'$, the bulk Lorentz factor $\Gamma$, and the magnetic dissipation power $P_\mathrm{diss}$ are shown in \cite{dgdp_2025} and \cite{Zhang2021}.
In this work, we adopt this reconnection-driven jet framework to model the emission of Mrk~501, constraining the corresponding jet parameters as follows.

We set the jet power equal to the Eddington luminosity of the central engine, $L_\mathrm{j} = L_\mathrm{Edd}(M_\mathrm{BH})$,
assuming a supermassive black hole mass of $M_\mathrm{BH} = 10^9~\mathrm{M_\odot}$, consistent with the estimate of \cite{Barth2002}.
Following the results of \cite{Takamoto2015, Guo2019, Kadowaki2021}, who measured reconnection rates in MHD simulations,
we adopt $\xi_\mathrm{rec} = 0.1$.
The stripe distribution index is set to $a = 3.5$, since, as noted in \cite{dgdp_2025}, the flow solutions are insensitive to this parameter when $a > 3$.
In contrast, the parameters $l_\mathrm{min}$ and $\Gamma_\infty$ are poorly constrained and are therefore treated as free parameters within the ranges
$l_\mathrm{min} \in [100, 1000]$ \cite{Giannios_2019} and $\Gamma_\infty \in [20, 50]$.

Building on the jet structure described above, we model the emission of Mrk~501 as the superposition of two zones located at different positions along the blazar jet: a \emph{base emission} region and a \emph{transient emission} region.
The base emission arises from particles accelerated by turbulent reconnection at the location where magnetic dissipation peaks, i.e., at $s_\mathrm{max}$ (see equation~\ref{smax}).
The transient emission region, in turn, is located upstream at $s_\mathrm{tr} \ll s_\mathrm{max}$, as sketched in Figure~\ref{fig:sketch}, where particles are likewise accelerated by turbulent reconnection. This zone is intermittent relative to the life-time of the base emission.
Being situated closer to the jet base, the transient region is smaller, more magnetised, and owning a lower bulk Lorentz factor than the base region.
We associate this transient zone with the  episodes of enhanced X-ray flux and the appearance of narrow TeV features in Mrk~501.

In both the base and transient zones, the energy distribution of accelerated electrons is described by a single power law with a squared exponential cut-off:
\begin{equation}
N_i = N_{0,i}
\left(
\frac{E_e}{E_{0,i}}
\right)^{-\alpha_i}
\exp\left[
-\left(\frac{E_e}{E_{\max,i}}\right)^2
\right],
\label{N_i}
\end{equation}
where $E_{0,i}$ and $E_{\max,i}$ are the minimum and maximum electron energies, $\alpha_i$ is the spectral index, and $N_{0,i}$ the normalisation constant. The subscript $i$ denotes the emission zone, with $i = \mathrm{b,tr}$ referring to the base and transient regions, respectively.

The distributions $N_i$ are normalised according to the local magnetic dissipation power:
\begin{equation}
\eta_i P_\mathrm{diss}(s_i) = \Gamma(s_i)^2 \beta(s_i) \pi c r_i^2 U'_i \, ,
\label{etaPdiss}
\end{equation}
where $\eta_i \ll 1$ is the fraction of $P_\mathrm{diss}$ transferred to non-thermal electrons, $\beta=(1-\Gamma^{-2})^{1/2}$, $r_i$ is the radius of the emitting volume, and
\begin{equation}
U'_i = \int_{E_{0,i}}^{\infty} dE_\mathrm{e} E_\mathrm{e} N_i(E_\mathrm{e}),
\label{u_i}
\end{equation}
is the co-moving energy density of the suprathermal electrons.

The radius of each emission zone is constrained by causality through the observed minimum variability timescale of $\Delta t_\mathrm{v,i}$. Thus, we parameterise the radii of the emitting regions as
\begin{equation}
r_i = f_{\mathrm{v},i} \frac{c \Delta t_\mathrm{v,i} \Gamma(s_i)}{(1+z)},
\label{r_i}
\end{equation}
where $f_{\mathrm{v},i} \leq 1$ is a dimensionless scaling factor and $z = 0.034$ is the redshift of the source.
The spectral index $\alpha_i$ and the minimum energy $E_{0,i}$ are treated as free parameters. In contrast, the maximum energy $E_{\max,i}$ is determined by equating the turbulent reconnection acceleration timescale with the corresponding radiative cooling timescale, as discussed in the following subsection. Consequently, turbulent reconnection leaves an imprint on the blazar SED through the spectral cut-offs associated with $E_{\max,i}$.
For each zone, $\Delta t_\mathrm{v,i}$ will be different since, by construction, the transient zone corresponds to a faster flux variability than the stable one. For instance, for the SEDs of Mrk~501 analysed in the next Section, we adopt a variability timescale of $\Delta t_{\mathrm{v,tr}}\approx 1$~day for the transient zone, motivated by the variations in the $\gamma$-ray flux during the analysed period \cite{MAGIC_2020}. 
On the other hand, we assume $\Delta t_{\mathrm{v,b}} \approx 3$~days for the stable zone, since the flux from optical to soft X-ray wavelengths appears relatively quiescent across the four SED datasets analysed.

Summarising, within the framework proposed here,  the particle energy distributions in both the base and transient emission zones are determined from equations~(\ref{N_i})–(\ref{r_i}) by specifying the model parameters $s_\mathrm{tr}$, $l_\mathrm{min}$, $\Gamma_\infty$, $E_{0,i}$, $\alpha_i$, $\eta_i$, and $f_{\mathrm{v},i}$.
These parameters are treated as free variables and are constrained through the modelling of the SED data.

\subsection{Particle 
 acceleration time}
\label{subsec:atime}
Particle acceleration in turbulent magnetic reconnection proceeds through a first-order Fermi process while particles remain confined within the reconnection region \cite{dgdp_2005,Lazarian_2012}. The particle's fractional energy gain is $\Delta E/E \propto v_{\rm rec}/c$, where $v_{\rm rec}$ is the reconnection velocity. During this process, the acceleration time is independent of the particle energy; this result has been confirmed by numerical simulations, which find only a very weak dependence ($t_\mathrm{acc} \propto E^{0.1-0.2}$) attributable to variations in $v_{\rm rec}$
\cite{del2016properties,liu2017does,medina2021particle,medina2023particle}. When the particle Larmor radius exceeds the thickness of the reconnection layer, particles can continue to gain energy by drifting along the reconnection layer \cite{kowal2012particle,Lazarian_2012,del2016properties, Zhang_2021, pino2024particle}, in which case the acceleration time increases approximately linearly with energy.

In the parameter space relevant to this study, radiative cooling prevents particle acceleration much before reaching the drift-acceleration regime. Hence, the maximum particle energy is determined by the Fermi-like acceleration phase. We estimate the corresponding acceleration time following the analytical results of \cite{Xu_2023}:
\begin{equation}
t_\mathrm{acc,F} \sim \frac{4\Delta}{cd_\mathrm{ur}},
\label{tacc_Xu}
\end{equation}
where $\Delta$ is the thickness of the acceleration region, here taken as the size of the emitting blob ($\Delta \sim 2r_i$), and
\begin{equation}
d_\mathrm{ur} = \frac{2\beta_\mathrm{rec}\left(3\beta_\mathrm{rec}^2 + 3\beta_\mathrm{rec} + 1\right)}{3\left(\frac{1}{2}+\beta_\mathrm{rec}\right)\left(1-\beta_\mathrm{rec}^2\right)},
\end{equation}
with $\beta_\mathrm{rec} = v_\mathrm{rec}/c$ and $v_\mathrm{rec}$ the local reconnection speed. As discussed in the previous section, we adopt $v_\mathrm{rec} = \xi_\mathrm{rec} v_\mathrm{A}$ with $\xi_\mathrm{rec} = 0.1$, and $v_\mathrm{A}$ the local Alfvén velocity.
Following \cite{dgdp_2025}, the Alfvén velocity at a given jet location $s$ is
\begin{equation}
v_\mathrm{A} = \frac{v_{\mathrm{A,0}}}{\sqrt{1+\left(v_{\mathrm{A,0}}/c\right)^2}},
\end{equation}
which allows for the relativistic and classical limits \cite{Somov_2006,Kadowaki_2015}, where $v_{\mathrm{A,0}}$ is the classical one:
\begin{equation}
v_{\mathrm{A,0}} = \frac{B'}{\sqrt{4\pi\rho'}} = c\sqrt{\frac{1-\chi}{\chi}}.
\label{v_A0}
\end{equation}
Here, $B'$ and $\rho'$ are the magnetic field and mass density in the co-moving frame, and $\chi = \Gamma(s)/\Gamma_\infty$ is a function of the jet location $s$, obtained from \cite{Giannios_2019} as the solution of
\begin{equation}
\frac{d\chi}{d\zeta} = \frac{(1-\chi)^{\kappa}}{\chi^2};\,\,\kappa = (3a-1)/(2a-2)
\end{equation}
with $\zeta \equiv 2\xi_\mathrm{rec}s/(l_\mathrm{min}\Gamma_\infty^2)$ (see \cite{Giannios_2019,dgdp_2025} for details).

The numerical simulations of test-particle acceleration 
within 3D MHD simulated reconnection in relativistic jets
performed by \cite{dgdp2024} show good agreement with the acceleration times predicted by equation~\ref{tacc_Xu}. The test particles in these simulations are protons; electrons are expected to behave similarly but with shorter timescales owing to their smaller Larmor radii. Therefore, in this work, which focuses on leptonic emission in blazar jets, we assume that electrons follow an acceleration time scaled as \cite{medina2021particle}:
\begin{equation}
t_\mathrm{acc,e} = \left(\frac{m_\mathrm{e}}{m_\mathrm{p}}\right) t_\mathrm{acc,F},
\label{t_acc_e}
\end{equation}
where $m_\mathrm{e}$ and $m_\mathrm{p}$ are
the electron and the proton masses, respectively.
\footnote{Because the Larmor radius is $r_{\rm L}=p_\perp c/(|q|B)$, particles with the same charge magnitude (electrons and protons) and comparable dynamical conditions in the same electromagnetic fields exhibit acceleration times that scale approximately with momentum and, for comparable Lorentz factors, with mass. Therefore, when extending the proton-based calibration of the Fermi acceleration time to leptons in the present leptonic-emission framework, we adopt the mass-scaled prescription (eq. \ref{t_acc_e}), as in \cite{medina2021particle}.}

\subsection{Maximum particle energy}

In our emission scenario, the particle energy distribution (equation~\ref{N_i}) is limited by the maximum energy attainable through turbulent reconnection acceleration. For both the base and transient zones, this maximum energy is obtained by equating the acceleration time (equation~\ref{t_acc_e}) with the total time-scale of particle energy losses.

In the reconnection region, the rate of electron diffusion is negligible compared to the rate of radiative losses. Thus, the maximum particle energy is determined by the condition
\begin{equation}
t_\mathrm{acc,e} = 
\left(
t^{-1}_\mathrm{syn}+t^{-1}_\mathrm{ssc}
\right)^{-1},
\end{equation}
where $t^{-1}_\mathrm{syn}$ and $t^{-1}_\mathrm{ssc}$ are the synchrotron and synchrotron self-Compton cooling rates, respectively. These rates depend on the co-moving magnetic field and particle energy and we calculate them as detailed in the  Appendix~A of \cite{dgdp_2025}.

\section{Application to Mkr 501}
\label{sec:application}

We apply the blazar reconnection framework outlined in the previous section to model the multiwavelength SEDs of Mkr~501 spanning from optical to VHE radiation. 
These SED datasets, taken from \cite{MAGIC_2020}, correspond to four consecutive days of observations, including the SED from the day preceding the TeV spectral spike (MJD~56957.98) and the SEDs obtained during the subsequent two days. 
In the following subsections, we describe the adopted modeling procedure and discuss the obtained outcomes.

\subsection{Spectral Energy Distributions}

For each SED dataset, the observed flux, $\nu F_\nu$, is modelled as the superposition of a transient (tr) component on top of a base (b) component,  
$\nu F_\nu = \nu F_{\nu,\mathrm{b}} + \nu F_{\nu,\mathrm{tr}}$.  
Each component includes synchrotron and SSC radiation from the non-thermal particle population.  
The differential flux at photon energy $E$ from zone $i$ (with $i=\mathrm{b},\mathrm{tr}$) is obtained as
\begin{equation}
\nu F_{\nu,i} = \frac{\Gamma_i^4}{4\pi D_\mathrm{L}^2}
\left(
\epsilon L^\mathrm{syn}_{\epsilon,i} +
\epsilon L^\mathrm{ssc}_{\epsilon,i}
\right),
\label{nuFnu_i}
\end{equation}
where $E = \Gamma_i \epsilon/(1+z)$, and where $\epsilon$, $L^\mathrm{syn}_\epsilon$, and $L^\mathrm{ssc}_\epsilon$ denote the photon energy and the differential synchrotron and SSC luminosities in the co-moving frame.  
These luminosities are computed following Appendix~B1 of \cite{dgdp_2025}, including internal $\gamma$--$\gamma$ absorption (relevant for the SSC component) and neglecting synchrotron self-absorption, which is negligible at the photon energies produced in this model.  
The formulation in equation~\ref{nuFnu_i} is valid for viewing angles $\theta_\mathrm{j}<1/\Gamma_\mathrm{j}$ relative to the jet propagation axis.

The synchrotron and SSC emission produced in each zone depends on the local macroscopic properties of the jet 
(such as $B'$, $\Gamma, v_\mathrm{A}$), 
as well as on the population of the accelerated particles, which
are determined through the jet–reconnection framework by specifying the input model parameters.

Based on observational constraints and theoretical results, we fix the following parameters in all calculated SEDs: the source redshift $z=0.034$, luminosity distance $D_\mathrm{L}=149.4$~Mpc, supermassive black hole mass $M_\mathrm{BH}=10^{9}\,\mathrm{M}_\odot$, reconnection rate $\xi_\mathrm{rec}=0.1$, and jet power $L_\mathrm{j}=L_\mathrm{Edd}(M_\mathrm{BH})$.
We then model the four-day SED sequence by varying the following free parameters, as described in Section \ref{sec:basic_model}:
\begin{itemize}
\item the jet terminal Lorentz factor, $\Gamma_\infty$,
\item the minimum stripe-length scaling factor, $f_\ell \equiv l_\mathrm{min}/R_\mathrm{g}$, with $R_\mathrm{g}$ being the gravitational
radius of the central BH,
\item the location of the transient emission zone, $s_\mathrm{tr}$,
\item the size-scaling factor of each emission zone, $f_{\mathrm{v},i} \equiv r_i(1+z)/(c\,\Delta t_\mathrm{v,i}\Gamma_i)$ (see equation~\ref{r_i}),
\item the fraction of magnetic dissipation power, $\eta_i$ (see equation~\ref{etaPdiss}),
\item the minimum energy of accelerated electrons, $E_{0,i}$, and
\item the spectral index of the particle distribution, $\alpha_i$ (see equation~\ref{N_i}),
\end{itemize}
resulting in a total of eleven free parameters (for $i=$b,tr).

To model the four days in a consistent manner, we proceed as follows.  
Once the parameters describing the jet structure ($\Gamma_\infty$, $f_\ell$), the base emission region ($f_\mathrm{v,b}$, $\eta_\mathrm{b}$, $E_{0,\mathrm{b}}$, $\alpha_\mathrm{b}$), and the location of the transient zone ($s_\mathrm{tr}$) are selected, they are kept fixed across all four days. This choice reflects that both emission zones belong to the same jet and that the base emission is assumed to remain quasi-stable over the four-day interval. We also assumed that the transient population of electrons has a similar $\alpha_\mathrm{tr}$, which produces the feature in the spectrum of 56857.98.
For each SED dataset, we therefore vary only the transient-zone parameters $f_\mathrm{v,tr}$, $\eta_\mathrm{tr}$, and $E_{0,\mathrm{tr}}$, 
to reproduce the observations of that specific day.

\begin{table}[b]
\centering
\caption{\label{tab:modelp}%
Input parameters for the blazar reconnection model described in Section~\ref{sec:basic_model}, which are kept fixed across the four analysed SEDs of  Figure \ref{fig:4SEDs}.
}
\label{table:free_params1}
\begin{tabular}{lcccc}
\hline \hline
\textrm{Parameter} & \textrm{Values}  \\
\hline
$\Gamma_\infty$          & 25.00  \\
$f_\ell$            & 100.00  \\
$f_\mathrm{v,b}$    & 0.15  \\
$\eta_\mathrm{b}$ $[10^{-3}]$         & 2.25  \\
$E_\mathrm{0,b}$ [$10^4 m_e c^2$]   & 1.62  \\ 
$\alpha_\mathrm{b}$ & 2.57   \\
$s_\mathrm{tr}$ [pc] & 0.35   \\
$\alpha_\mathrm{tr}$ & 2.80  \\
\hline \hline
\end{tabular}
\end{table}

\begin{table}[b]
\caption{\label{tab:modelp}%
Input parameters for the blazar reconnection model described in Section~\ref{sec:basic_model}, which change from one SED dataset to another and are associated with different observing days.
MJD 56, MJD 57, MJD 58, and MJD 59 are abbreviations for MJD 56856.91, MJD 56857.98, MJD 56858.98, and MJD 56859.97, respectively.}
\label{table:free_params2}
\begin{tabular}{lcccc}
\hline \hline
\textrm{Parameter} &
\multicolumn{1}{c}{\textrm{MJD 56}}&
\multicolumn{1}{c}{\textrm{MJD 57}}&
\multicolumn{1}{c}{\textrm{MJD 58}} & 
\multicolumn{1}{c}{\textrm{MJD 59}}\\
\hline
$f_\mathrm{v,tr} [10^{-3}]$  & 1.5  & 2.7  & 1.9 & 0.6\\

$\eta_\mathrm{tr} [10^{-3}]$ & 2.1  & 2.5  & 2.3 & 0.1 \\

$E_\mathrm{0,tr}$ [$10^5 m_\mathrm{e}c^2$] & 5.0  & 5.8  & 4.2 & 4.8\\
\hline \hline
\end{tabular}
\end{table}

\begin{figure*}
   \centering
   \includegraphics[width=\hsize]{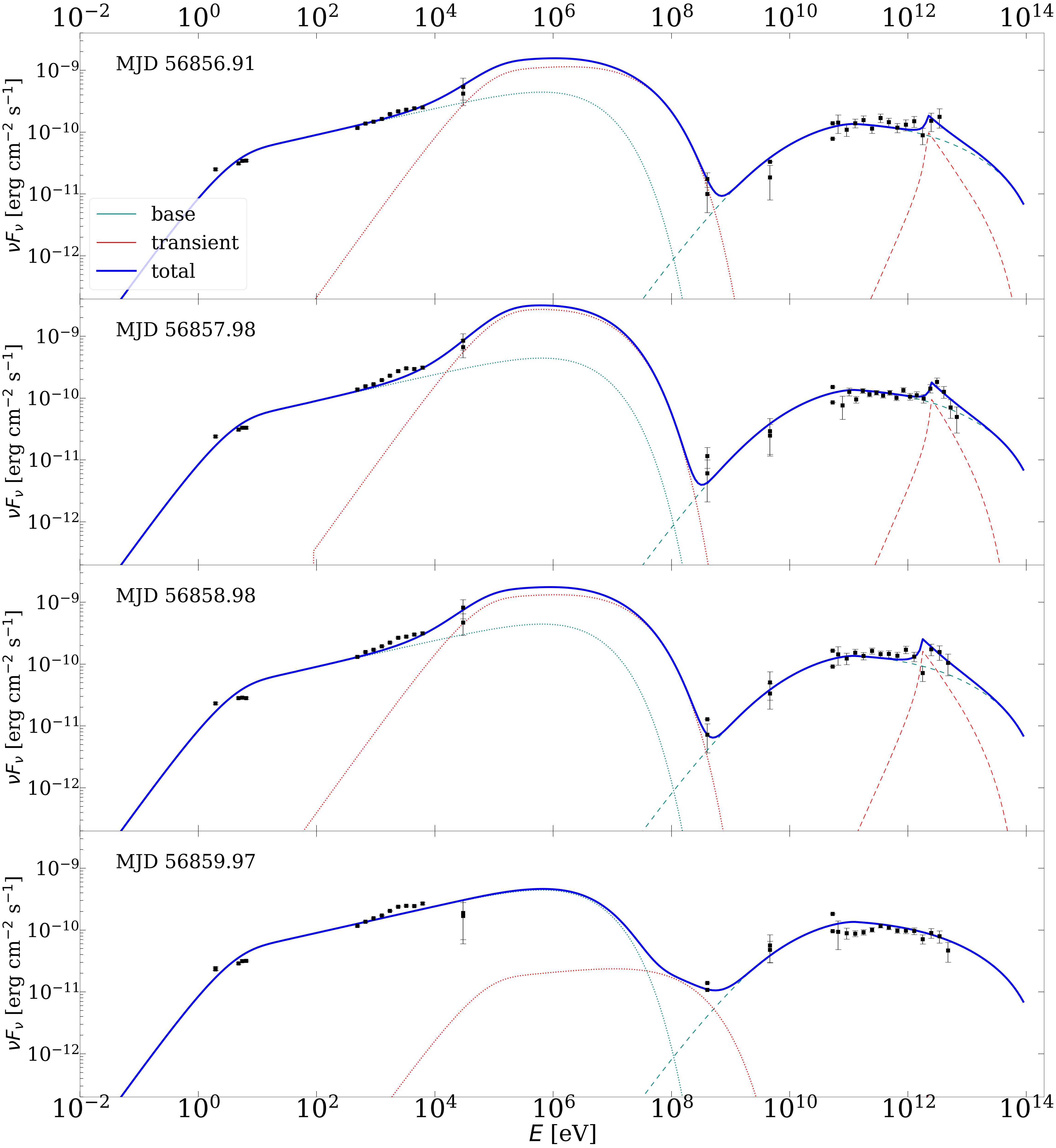}
\caption{
Multi-wavelength SEDs of Mrk 501 measured on four different days selected from the daily datasets reported in \cite{MAGIC_2020}. The overlaid curves are obtained through the two-zone, turbulent-driven reconnection acceleration jet model discussed in the text.
Synchrotron and SSC emissions are represented by dotted and dashed curve styles, respectively. Curves shown in different colours corresponds to the different considered zones, as indicated.}
        \label{fig:4SEDs}
\end{figure*}

In Fig.~\ref{fig:4SEDs}, we present the four-day SEDs resulting from the modelling procedure described above. Each frame in this figure corresponds to a given observation day, progressing from the top to the bottom panels. The model parameters adopted to produce these SEDs are listed in Tables ~\ref{table:free_params1} and ~\ref{table:free_params2}
 , with the first being related to the unchanged parameters, and the second to the parameters that vary each observational day in the transient zone.
 
As shown in Fig.~\ref{fig:4SEDs}, the quasi-constant behavior of the spectral points comprising the first hump from optical to soft X-rays is generally well represented by the stable zone component of the model. In contrast, the evolution of the spectral spike in the associated model requires varying the parameters of the transient zone for each SED dataset. Under this interpretation, the rise and  the fall of the highest-energy X-ray spectral point, accompanied by the rise and fall of the spectral spike in the TeV range, can be attributed to the activation/de-activation of the transient zone in the proposed model.

\subsection{Acceleration and cooling times}

\begin{figure}
   \centering
   \includegraphics[width=6 cm]{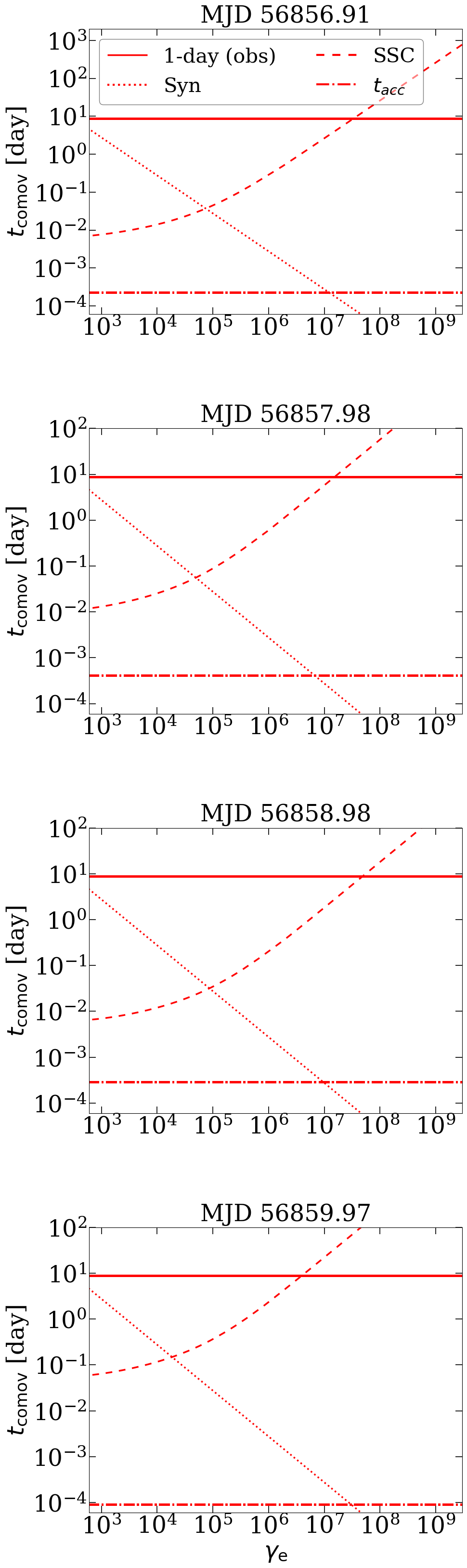}
    \caption{
Acceleration time (dotted-dashed lines) and cooling time of synchrotron (dotted lines)  and SSC (dashed lines) radiation for the transient emission region of the SED models of Figure~\ref{fig:4SEDs}. The horizontal solid lines indicate the one-day interval among the SED datasets, as seen in co-moving frame of the transient emission zone. 
}          
\label{fig:4cooling}
\end{figure}

The cooling times, compared with the Fermi acceleration timescales driven by turbulence-induced reconnection, for each day, are shown in Figure \ref{fig:4cooling}.

The Fermi acceleration timescales, and synchrotron plus SSC cooling are within the adopted variability timescale (indicated by the thick lines, already transformed into the reference frames of each zone) for both the base and transient regions.
We convert the observed variability timescale to the comoving frame of each emission zone as
$\Delta t_{\mathrm{comov},i} = \Gamma_i (1+z)\, \Delta t_{\mathrm{v},i}$,
where $\Gamma_i$ is the Lorentz factor of zone $i$, and $\Delta t_{\mathrm{v},i}$ is the observed variability timescale
(respectively, 1-day for the transient zone and 3-days for the stable zone). In the base zone, this occurs nearly after the radiation is emitted predominantly through the SSC mechanism.

\begin{figure}
   \centering
   \includegraphics[width=8 cm]{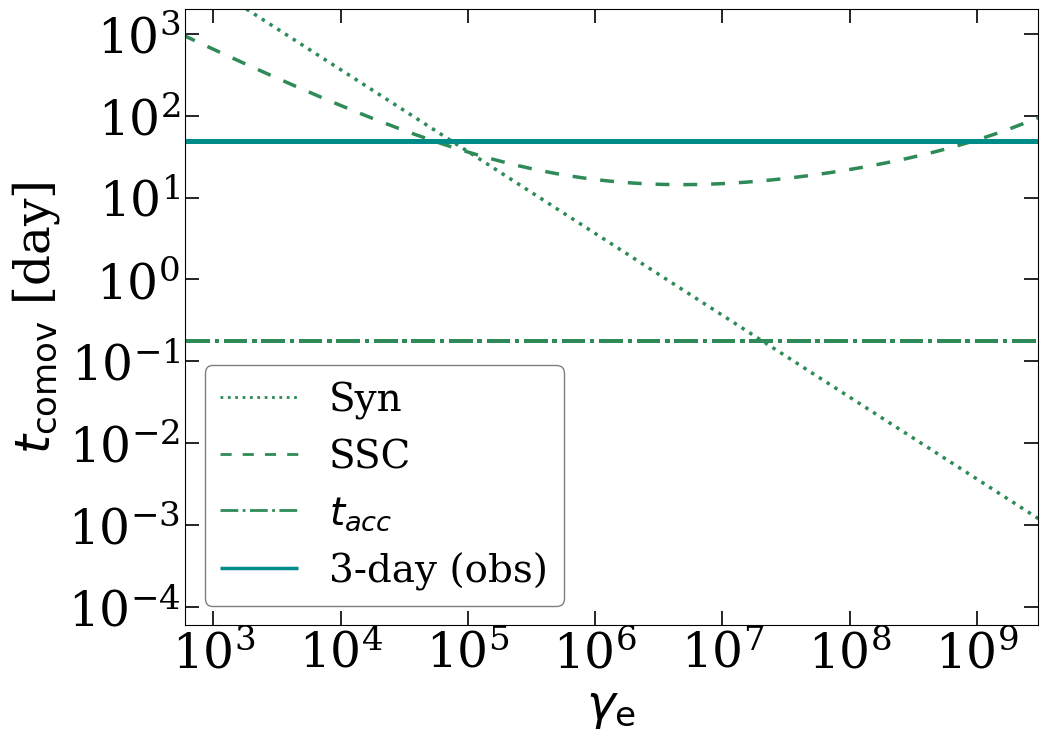}
    \caption{
Acceleration time (dotted-dashed line) and cooling time of synchrotron (dotted line)  and SSC (dashed line) radiation for the stable mission region of the SED models of Figure~\ref{fig:4SEDs}. The horizontal solid line indicates the three-day interval, as seen in co-moving frame of the base emission zone. }          
\label{fig:5cooling}
\end{figure}

\begin{table}[b]
\centering
\caption{\label{tab:model2}%
Output parameters for the base and transient zones of the jet emission model discussed in text, which are common to all four analysed SEDs of Figure \ref{fig:4SEDs}.}
\label{table:output1}
\begin{tabular}{lc}
\hline \hline
\textrm{Parameter} \footnote{``\textbf{lg} $x$'' indicates the base 10 logarithm of the quantity $x$.}&
\textrm{Values} \\
\hline
$s_\mathrm{b}=
s_\mathrm{max}$ [pc]          & 5.005 \\
$\Gamma_{\mathrm{b}}$ & 16.049 \\
$\Gamma_{\mathrm{tr}}$ & 8.540 \\
$B'_{\mathrm{b}} [\mathrm{G}]$ & 0.049 \\
$B'_{\mathrm{tr}} [\mathrm{G}]$ & 1.803 \\
$(L_B / L_K)_{\mathrm{b}}$   &   0.558 \\
$(L_B / L_K)_{\mathrm{tr}}$   &   1.928 \\
lg $u_\mathrm{e,\mathrm{b}}$ $[\mathrm{erg/cm^3}]$ & -2.358\\
 lg $R_{\mathrm{b}}$ [cm]& 16.258 \\
 $E_\mathrm{max,b}$ [$10^7 m_e c^2$]   & 2.05 \\
\hline \hline
\end{tabular}
\end{table}

\begin{table}[h]
\caption{\label{tab:modelp}.
Output parameters for the 'transient' zone that vary in our jet emission model across the four analysed SEDs of Figure \ref{fig:4SEDs}, as discussed in the text. MJD 56, MJD 57, MJD 58, and MJD 59 are abbreviations for MJD 56856.91, MJD 56857.98, MJD 56858.98, and MJD 56859.97, respectively.
}
\label{table:output2}
\begin{tabular}{lcccc}
\hline \hline
\textrm{Parameter} \footnote{``\textbf{lg} $x$'' indicates the base 10 logarithm of the quantity $x$.}&
\multicolumn{1}{c}{\textrm{MJD 56}}&
\multicolumn{1}{c}{\textrm{MJD 57}}&
\multicolumn{1}{c}{\textrm{MJD 58}} & \multicolumn{1}{c}{\textrm{MJD 59}}\\
\hline
lg $u_\mathrm{e,\mathrm{tr}}$  $[\mathrm{erg/cm^3}]$ & 3.540 & 3.105 & 3.374  & 3.014 \\
lg $R_{\mathrm{tr}} [\mathrm{cm}]$ & 13.507 & 13.762 & 13.609 &  13.109 \\
$E_\mathrm{max,tr}$ [$10^6 m_\mathrm{e} c^2$]   & 12.078  & 6.710 & 9.536 & 30.196\\
\hline \hline
\end{tabular}
\end{table}

\subsection{Properties of the emitting regions}

\begin{figure}
   \centering
   \includegraphics[width=\hsize]{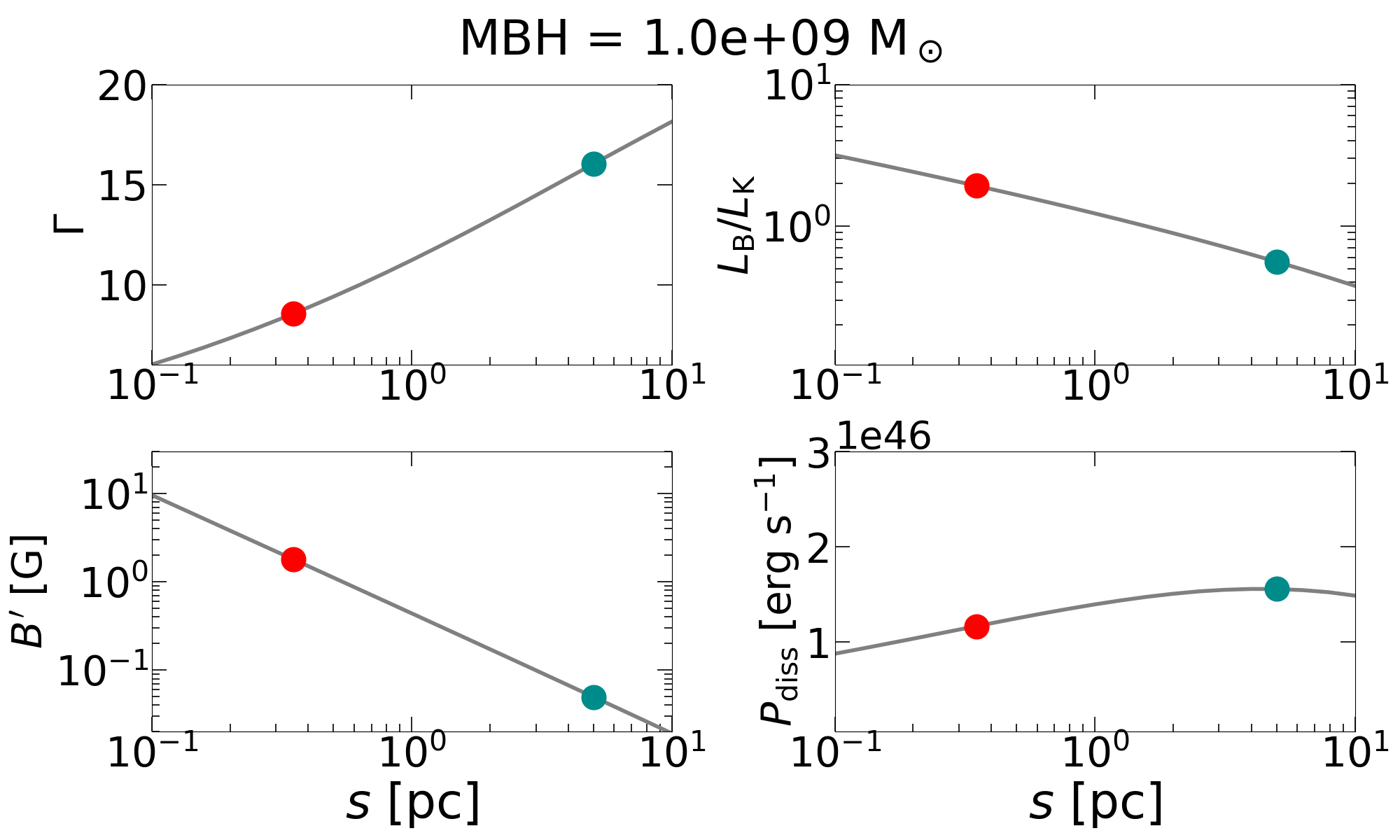}
\caption{
Jet properties predicted by the striped jet model 
using the input parameters specified in the text and in Table~\ref{table:free_params1}. The coloured points indicate the locations of each emission zone adopted for computing the SED models shown in Figure~\ref{fig:4SEDs} (red for the transient zone; sea green for the base zone). $\Gamma$  is the Lorentz factor, transient), $L_B/L_K$ is the magnetic/kinetic power ratio, $B'$ is the co-moving magnetic field, and $P_{\mathrm{diss}}$ is the magnetic dissipation power.}
         \label{fig:jetprops}
\end{figure}

Figure \ref{fig:jetprops} summarizes  the jet properties predicted by the striped-jet solution for the input parameters adopted in this work, and marks the locations of the transient and base emission zones used to compute the SED sequence.
The corresponding output parameters are provided in Table \ref{table:output1} (constant terms) and Table \ref{table:output2} (variable terms).

As expected for a gradually accelerating, dissipative outflow, the bulk Lorentz factor increases with distance along the jet. Quantitatively, we obtain $\Gamma_{\rm tr}\simeq 8.54$ in the transient zone and $\Gamma_b\simeq 16.05$ in the base zone (Table  \ref{table:output1}). Over the same interval, the comoving magnetic field decreases steeply, from $B'_{\rm tr}\simeq 1.80~{\rm G}$ to $B'_b\simeq 4.9\times 10^{-2}~{\rm G}$ (Table  \ref{table:output1}). This strong decline in $B'$ is also reflected in the magnetic/kinetic power ratio, which drops from $(L_B/L_K)_{\rm tr}\simeq 1.93$ to $(L_B/L_K)_b\simeq 0.56$ (Table  \ref{table:output1}). Thus, the transient region is substantially more magnetized than the base region, consistent with being located closer to the jet launching/dissipation onset.

In contrast to the pronounced evolution of $\Gamma$ and $B'$, the magnetic dissipation power $P_{\mathrm{diss}}$ varies only mildly between the two emitting locations (Figure \ref{fig:jetprops}), so that the two zones can radiate comparable total dissipated power while exhibiting markedly different local conditions. This is particularly important for the narrow TeV feature: reproducing the spike and the contemporaneous hard X-ray enhancement requires a compact, strongly magnetized transient region with a high density of radiating particles.

Indeed, the inferred electron energy density differs dramatically between the two zones. The base region has $\log u_{e,b}\simeq -2.36$ (in ${\rm erg~cm^{-3}}$), whereas the transient zone spans $\log u_{e,{\rm tr}}\simeq 3.01$--$3.54$ across the four SEDs (Tables  \ref{table:output1} and \ref{table:output2}), i.e. a contrast of $\sim 10^{5}$. In the model, this day-to-day variability in $u_{e,{\rm tr}}$ tracks the changing strength of the transient component and, in particular, the prominence of the $\sim$TeV spike.

The characteristic sizes of the emission regions follow from the adopted variability constraints. The base zone radius is $R_b\simeq 10^{16.26}~{\rm cm}$ (Table \ref{table:output1}), while the transient zone is much smaller, with $R_{\rm tr}\simeq 10^{13.11}$--$10^{13.76}~{\rm cm}$ (Table \ref{table:output1}), consistent with faster variability and with the need for a compact region to produce a narrow spectral component. The transient radius varies only weakly from day to day compared to the changes in $u_{e,{\rm tr}}$.

Finally, the maximum energies implied by balancing turbulent-reconnection acceleration and radiative losses differ between zones and evolve across the SED sequence. The base region reaches $E_{\max,b}\simeq 2.05\times 10^{7}\,m_ec^{2}$ (Table \ref{table:output1}). In the transient region, $E_{\max,{\rm tr}}$ varies between $\simeq (6.7$--$30)\times 10^{6}\,m_ec^{2}$ depending on the observing day (Table \ref{table:output2}), indicating that modest changes in the transient-zone conditions can shift the high-energy cutoff and, consequently, reshape both the hard X-ray synchrotron tail and the TeV SSC output.

\subsection{Investigation of the parametric space}

\begin{figure*}
   \centering
   \includegraphics[width=0.95\linewidth]{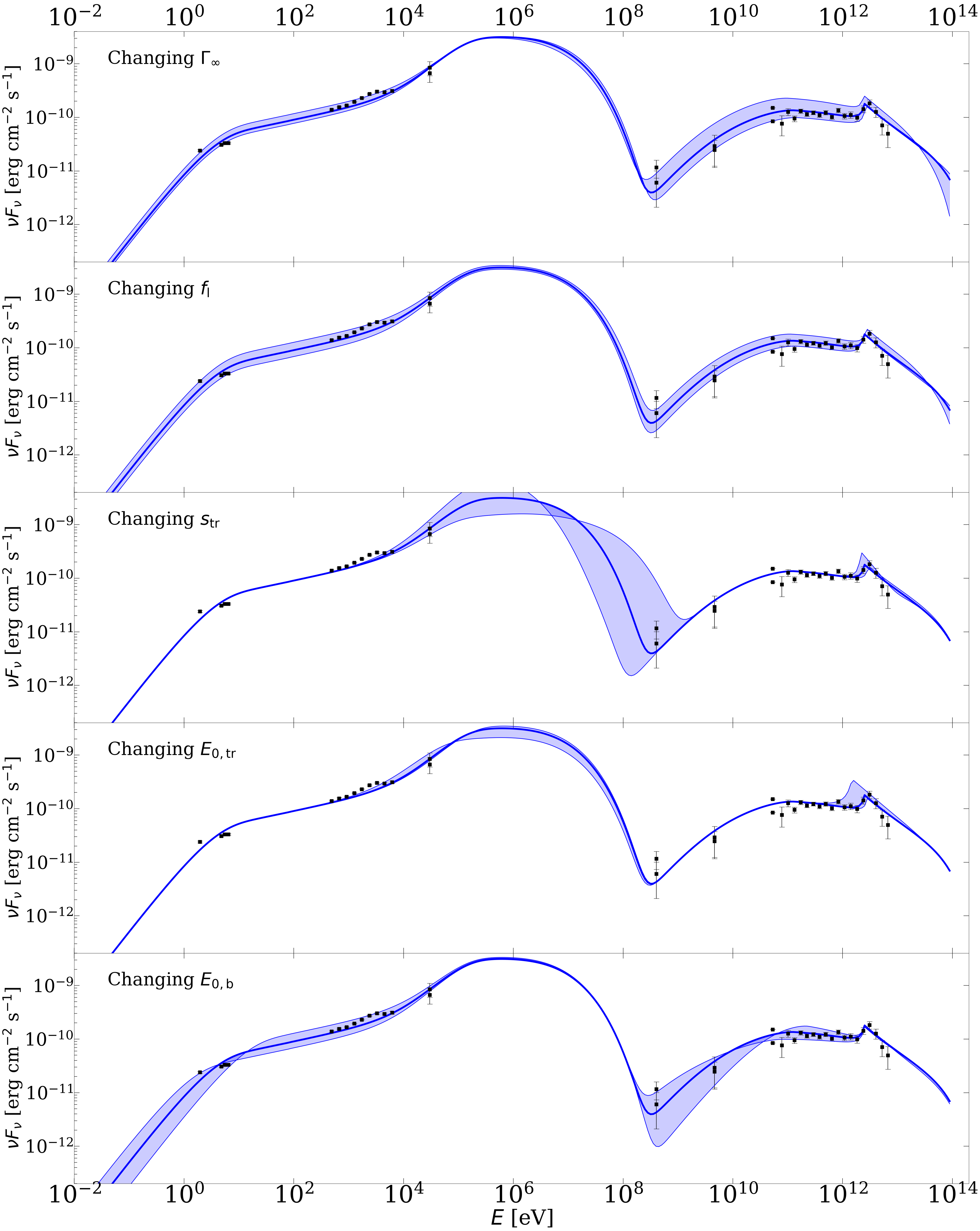}
\caption{Parameter study of the multi-wavelength SED of Mrk 501 at MJD 56857.98 reported in \cite{MAGIC_2020}. 
In each plot, a single parameter was varied relative the reference model shown in third panel of Figure \ref{fig:4SEDs} (solid blue line), while all other parameters were kept fixed, following Tables \ref{table:free_params1} and \ref{table:free_params2}. From top to bottom, the plots show the following variations:  
$\Gamma_{\infty}=[22,27]$;$f_{\mathrm{l}}=[80,120]$; $s_{\mathrm{tr}}=[0.20,0.60]$;$E_{\mathrm{0,tr}}=[3.5,6.5] \times10^5 m_{\mathrm{e}}c^2$;  
$E_{\mathrm{0,b}}=[1.0,3.0] \times10^4 m_{\mathrm{e}}c^2$.
The blue shaded regions represent the outcome in the total flux of using the selected ranges.} 
        \label{fig:4change1}
\end{figure*}

\begin{figure*}
   \centering
   \includegraphics[width=\hsize]{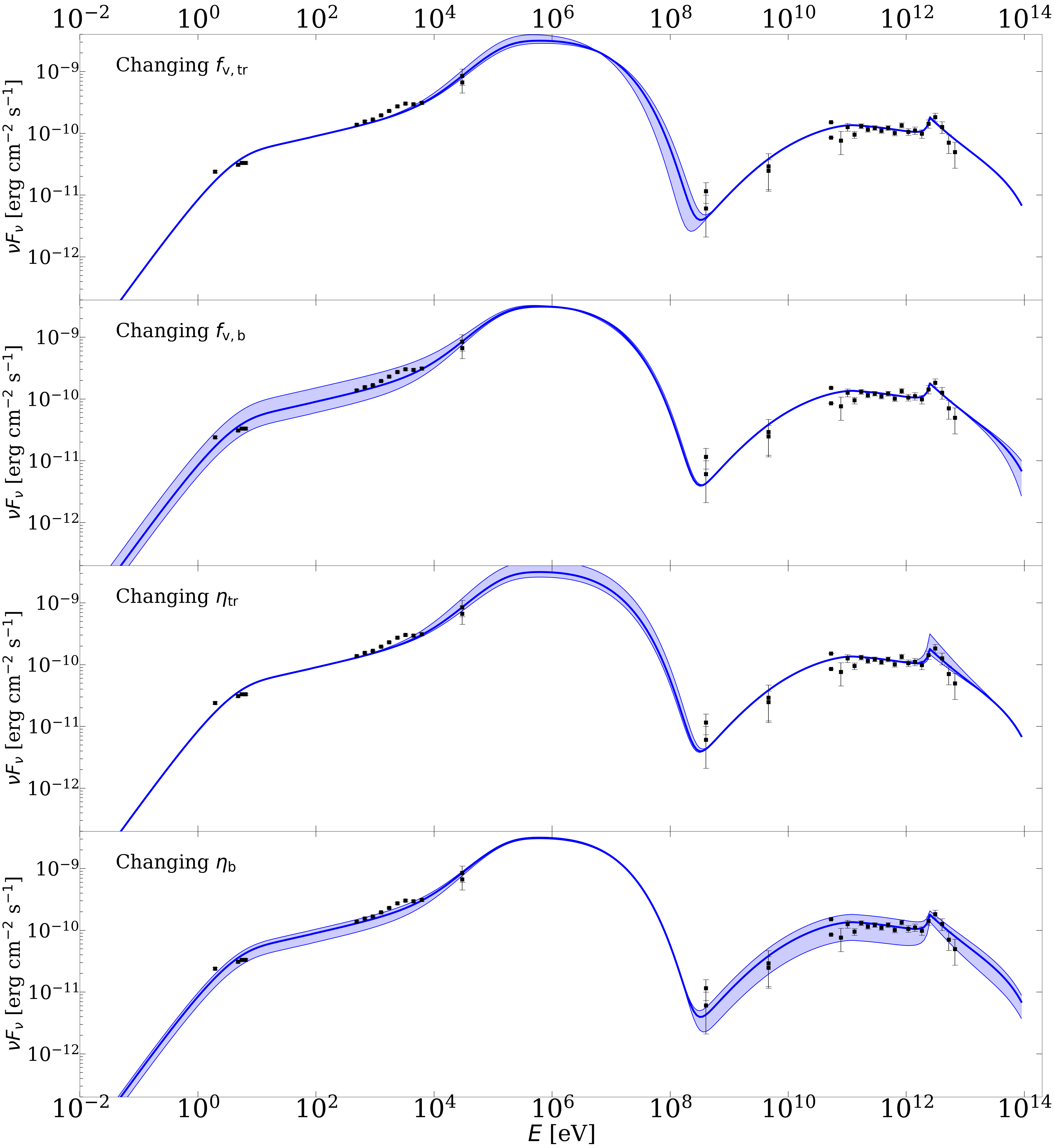}
\caption{
Same as in Figure~\ref{fig:4change1}, but showing the outcome of the following variations: $f_{\mathrm{v,tr}}=[2.4,3.5] \times 10^{-3}$; $f_{\mathrm{v,b}}=[0.10,0.25]$; $\eta_{\mathrm{tr}}=[2.0,4.0] \times 10^{-3}$; $\eta_{\mathrm{b}}=[1.6,2.6] \times 10^{-3}$.}
        \label{fig:4change2}
\end{figure*}

\begin{figure*}
   \centering
   \includegraphics[width=\hsize]{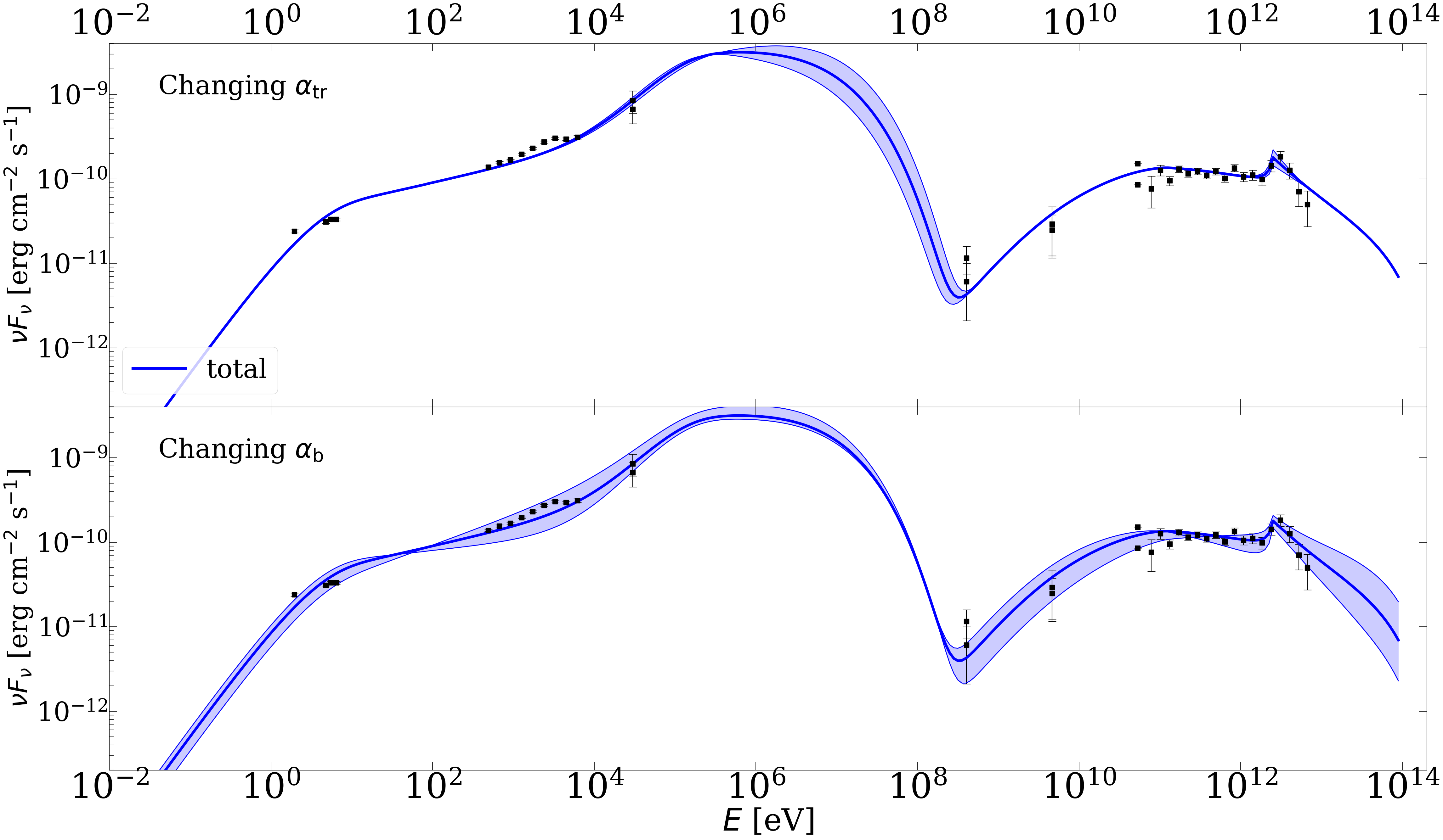}
\caption{
Same as in Figure~\ref{fig:4change1}, but showing the outcome of varying within the parameter ranges
$\alpha_{\mathrm{tr}}=[2.4,3.2]$; $\alpha_{\mathrm{b}}=[2.3,2.8]$.
}
        \label{fig:4change3}
\end{figure*}

Figure \ref{fig:4SEDs} presents our 
reference model for the set of free parameters (reported in   Tables~\ref{table:free_params1} and~\ref{table:free_params2}), simultaneously matching the four-day data sets. Although this fiducial setup is adopted as the reference solution, no formal spectral-fitting procedure was carried out, and the parameter values are therefore not uniquely constrained, permitting some variation within physically plausible ranges for this source. To demonstrate this, the SEDs in  Figures \ref{fig:4change1}, \ref{fig:4change2}, and \ref{fig:4change3} show how the results change for the SED of Mrk 501  at MJD 56857.98, when selected free parameters are varied around their fiducial values listed in  Tables~\ref{table:free_params1} and~\ref{table:free_params2}.

The range of parameters chosen were the following: $\Gamma_{\infty}=[22,27]$;$f_{\mathrm{l}}=[80,120]$; $s_{\mathrm{tr}}=[0.20,0.60]$ pc;
$E_{\mathrm{0,tr}}=[3.5,6.5] \times10^5 m_{\mathrm{e}}c^2$;
$E_{\mathrm{0,b}}=[1.0,3.0] \times10^4 m_{\mathrm{e}}c^2$;
$f_{\mathrm{v,tr}}=[2.4,3.5] \times 10^{-3}$; $f_{\mathrm{v,b}}=[0.10,0.25]$; $\eta_{\mathrm{tr}}=[2.0,4.0] \times 10^{-3}$; $\eta_{\mathrm{b}}=[1.6,2.6] \times 10^{-3}$; $\alpha_{\mathrm{tr}}=[2.4,3.2]$; $\alpha_{\mathrm{b}}=[2.3,2.8]$.

Figure \ref{fig:4change1} shows that variations in $\Gamma_{\infty}$ produce only a modest change in the flux of the first spectral hump, but a much stronger effect on the second hump. For $\Gamma_{\infty}=22$, the model underpredicts the spectral points in the 0.01--1 TeV range. In contrast, for $\Gamma_{\infty}=27$, the model overpredicts the observed flux in the same energy range.

The parameter $f_l$, on the other hand,   controls the overall spectral flux.  Varying it within the range $[80,120]$ slightly modifies the  spectrum, producing total curves that generally lie, respectively, below and above the reference model.

The parameter $E_{\mathrm{0,b}}$ controls the minimum spectral range and therefore affects the entire curve produced by the base zone. Setting $E_{\mathrm{0,b}} = 10^4 m_e c^2$ broadens the overall spectrum, which lowers the flux in the low-energy X-ray band and in the VHE gamma-ray range before the spectral spike. By contrast, increasing this parameter by a factor of three compresses the spectrum over the same energy ranges, raising the flux above the observed data.

Parameters $s_{\mathrm{tr}}$ and $E_{0,\mathrm{tr}}$ are the main ones governing the spectral shape in the transient zone. Both can strongly affect the sharpness of the spectrum, the position of the synchrotron peak, and the appearance and position of TeV spike.
$s_{\mathrm{tr}}$, in particular,  affects the first hump produced in the transient zone. When the acceleration/emission region is located closer to the SMBH, as in the case $s_{\mathrm{tr}}=0.20$, both the first hump ($\sim 10^5$ eV) and the spectral spike become softer, leading to a poorer agreement with the spectral points in the GeV range. On the other hand, for $s_{\mathrm{tr}}=0.60$, the first hump becomes much steeper, again showing disagreement with the spectral points in the same energy range.
The minimum and maximum range of values for $E_{0,\mathrm{tr}}$ produce, respectively, the broadening and narrowing, and the softening and hardening, of the spectral spike. As can be seen from Figure \ref{fig:4change1}, a value of $3.5 \times 10^5 m_\mathrm{e}c^2$ can shift the peak of the spectral component to $\sim$1 TeV, which differs from what was measured at $\sim$3 TeV.

In Figure \ref{fig:4change2}, both  $f_{\mathrm{v,tr}}$ and $f_{\mathrm{v,b}}$ parameters  affect only  the first hump.
In particular, a smaller value of $f_{\mathrm{v,b}}$, such as $0.10$, produces a first hump in the base zone that lies below the observed flux, whereas a larger value, such as $0.25$, leads to the opposite behavior.

$\eta_{\mathrm{tr}}$ and $\eta_{\mathrm{b}}$ control the overall flux level of each emitting zone. The selected range of values for both parameters provides a good representation specially  of the transient emission region, as shown in Figure \ref{fig:4change2}.

Finally, the spectral indices of the electron distributions, $\alpha_{\mathrm{tr}}$ and $\alpha_{\mathrm{b}}$, can modify the shape of the spectral tail in the transient and base zones. As shown in Figure \ref{fig:4change3}, the transient component is more degenerate, with a slightly sharper spectral spike for a softer spectral index ($\alpha_{\mathrm{tr}}=3.2$) than in our reference model ($\alpha_{\mathrm{tr}}=2.8$). 
In contrast, varying the base component shows that values as hard as 2.3 lead to a hardening of the spectral curve, resulting in a poorer match to the high-energy spectral points in the VHE range, which appear flatter.

\subsection{Further constraining the parameter space with $\chi^2$ minimization }

In the previous section, we adopted an intuitive approach by varying individual free parameters within a fiducial range around the SED reference model of Mrk 501 at MJD 56857.98. Here, we present a more refined parameter analysis based on a $\chi^2$ minimization method.
Specifically, we minimize a {\it joint} function of the form
$\mathcal{F}({\bf \Theta}) = \sum w_i\chi_i^2$,
which combines the $\chi^2$ tests of each 
dataset, to achieve the overall best fit for the four days, where 
$w_i$ is 
the weight assigned to the dataset $i$. We refer to ~\ref{appendix_optim} for further details on this multi-dataset fitting procedure.

\begin{figure*}[h!]
\centering

\begin{subfigure}{0.8\textwidth}   
    \centering
    \includegraphics[width=0.9\textwidth]{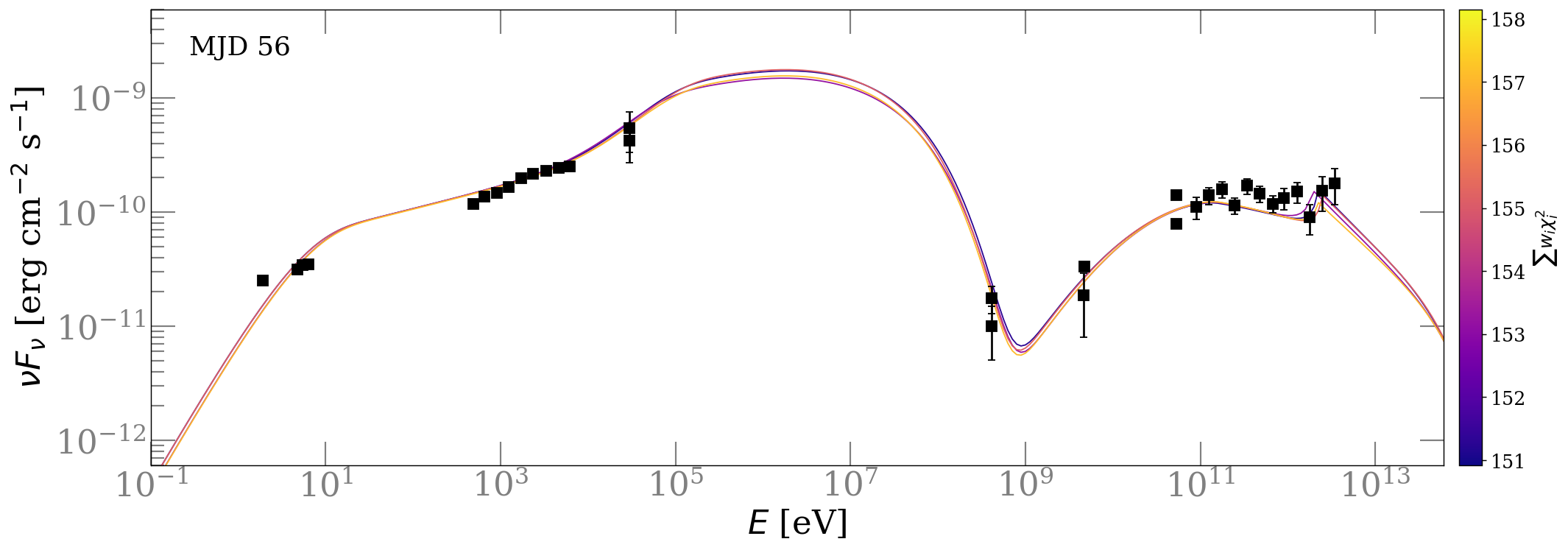}
\end{subfigure}

\vspace{0.0cm}

\begin{subfigure}{0.8\textwidth}   
    \centering
    \includegraphics[width=0.9\textwidth]{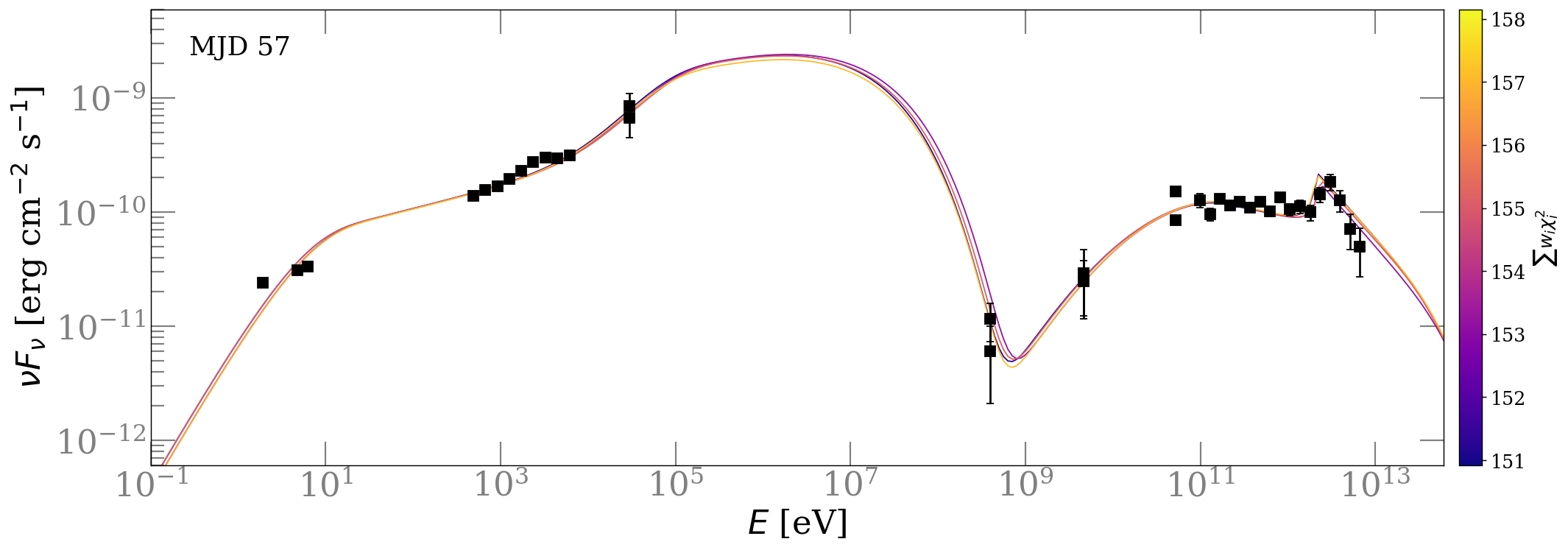}
\end{subfigure}

\vspace{0.0cm}

\begin{subfigure}{0.8\textwidth}   
    \centering
    \includegraphics[width=0.9\textwidth]{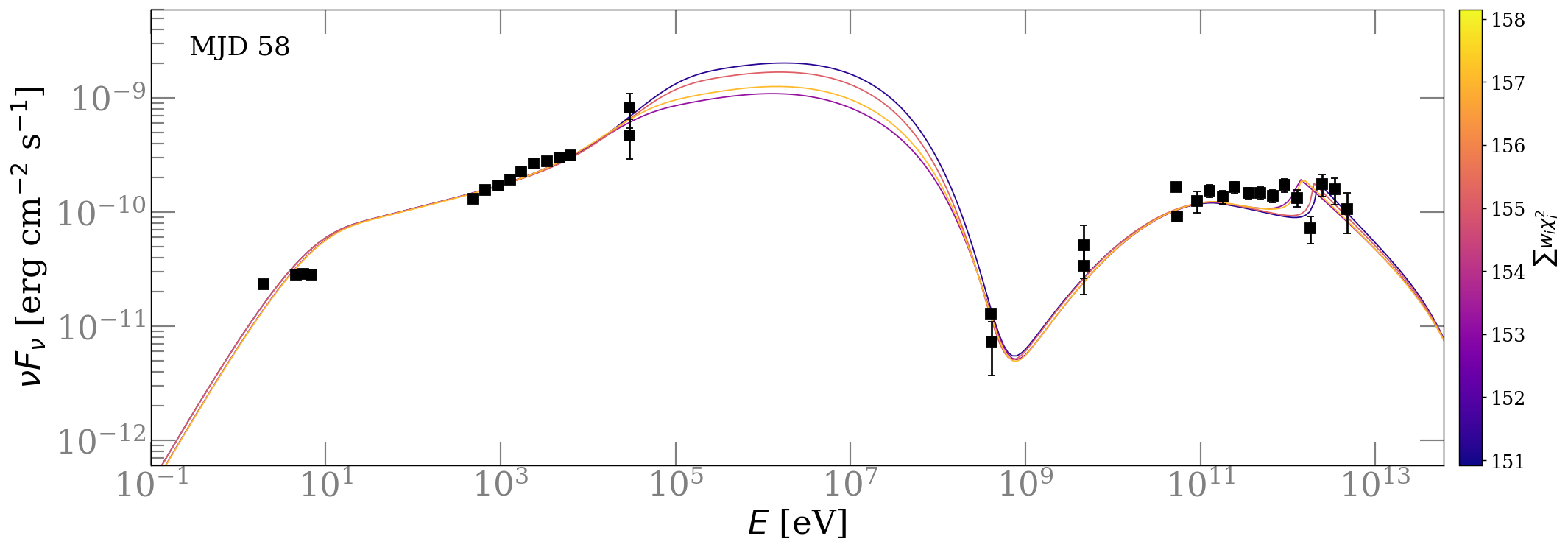}
\end{subfigure}

\vspace{0.0cm}

\begin{subfigure}{0.8\textwidth}   
    \centering
    \includegraphics[width=0.9\textwidth]{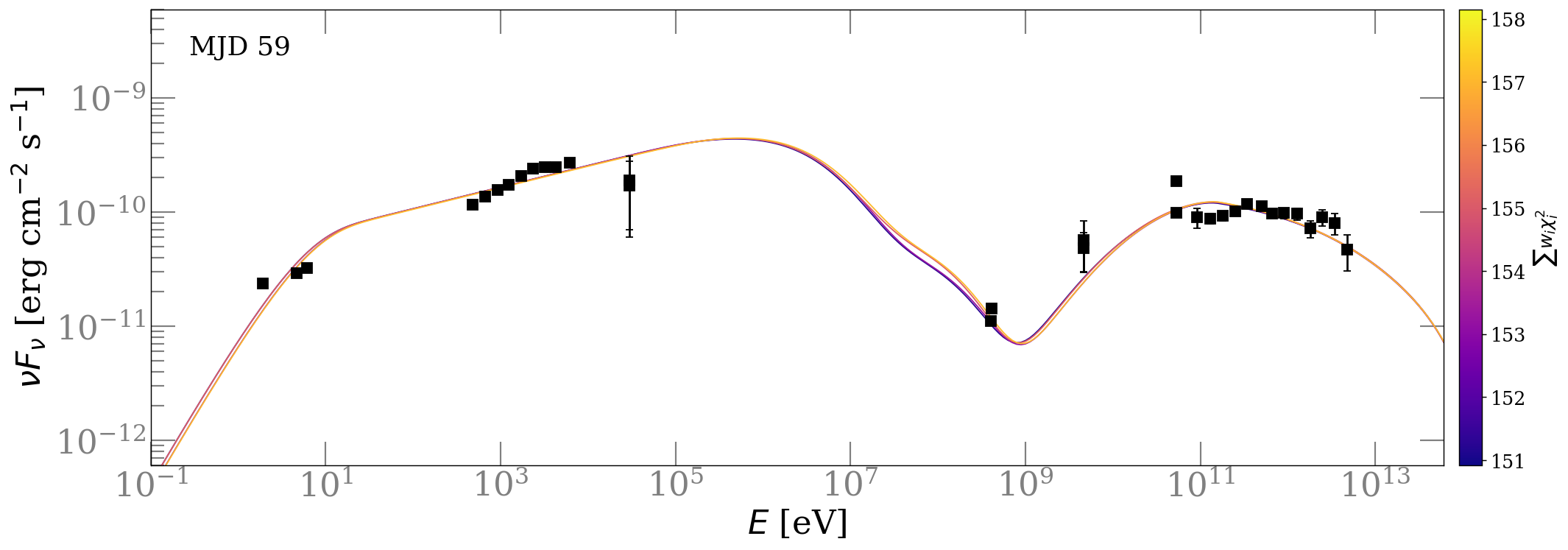}
\end{subfigure}

\caption{Optimized $(\chi^2$)fits of the multi-wavelength SEDs of Mrk 501 measured on four different selected days. Each colored total curve corresponds to the resulting value of the joint merit function, following the legend shown on the right.}
\label{fig:sedsopt}
\end{figure*}

Figure \ref{fig:sedsopt} shows the most consistent fits obtained for the respective datasets. Tables~\ref{tab:globalopt} and~\ref{tab:localopt} present the global and local parameters derived from the four best runs. The values of $\mathcal{F}({\bf \Theta})$ are also shown in Table~\ref{tab:globalopt}. 

Interestingly, for the global parameters, $E_{0,\mathrm{b}}$ shows the highest discrepancy with respect to the reference model (see Tables~\ref{table:free_params1}). Averaging $E_{0,\mathrm{b}}$ over the four runs yields a difference of $\sim 0.42\,m_e c^2$ with its original value. 

Analyzing the local parameters, $f_{v,\mathrm{tr}}$ and $E_{0,\mathrm{tr}}$ also have 
slight discrepancies when examining each run and each dataset separately. Specifically, for MJD 56857.98 and MJD 56859.97, these parameters deviate more from their counterparts in Table~\ref{table:free_params2} than those of the other datasets. For these two days of observational data, we assigned higher weights (see \ref{appendix_optim}).

\section{Summary and discussion}

The blazar Mrk 501 exhibited a distinct spike in its SED on July 19, 2014, as detected by the MAGIC telescopes during a multi-wavelength campaign. This event was accompanied by a marked increase in flux and a hardening of the X-ray spectrum, as observed by Swift-XRT.

To explain this spectral feature, we employed a two-zone-jet-emission scenario, in which the macroscopic properties of the emission regions follow the striped-jet model of \cite{Giannios_2019}. In addition, we assumed that turbulence-driven magnetic reconnection is responsible for electron acceleration, with energy dissipation occurring within reconnection layers. 
This approach is motivated by previous studies that demonstrated efficient Fermi proton acceleration beyond PeV energies, by tracking 
test particles in 3D MHD simulations of jets \citep{medina2021particle,medina2023particle,pino2024particle}, as well as by its successful application to reproducing the multi-messenger SED of the blazar TXS~0501+256 \citep{dgdp_2025}.
In the present framework, however, we  restricted our treatment to emission produced by electrons. Their  acceleration timescale, 
which is also attributed to turbulent reconnection-driven Fermi acceleration is given by eq. \ref{t_acc_e}.

In the proposed scenario, the regular emission of the source is produced at the peak of magnetic dissipation power in the jet, which we identify as the stable emission component. Superimposed on this, a second emission region of transient nature and located upstream in the jet, produces the spike observed in the VHE SED as well as the hardening of the observed X-ray emission. Under this hypothesis, the SED model profile is determined by eleven free parameters, as described in Section~\ref{sec:application}.

Within this framework, the model successfully reproduces the
TeV spectral spike observed on July~19th, while remaining consistent with the observed SEDs during the days immediately before and after this peculiar feature.  And, in particular, the appearance and disappearance of the spike, together with the associated X-ray hardening over the four analysed days, are explained by a short-term dissipation episode in the transient emission region (see Figure~\ref{fig:4SEDs}).

From the parameters associated with the transient zone, we
find a progressive increase in the causality factor, the fraction of magnetic dissipation power, and the minimum electron energy of the transient zone up to the day of the spectral spike (MJD 56857.98). After this peak, these parameters decrease on the following day, tracking the gradual weakening of the transient contribution to the spectra. Apart from the minimum electron energy, the other parameters show a decay by MJD 56859.97, which correlates with the disappearance of the spectral spike from the VHE range.

According to the present results,
the maximum energy attained by the accelerated electrons in the transient zone,
$E_{\mathrm{max,tr}}$,
produces variable synchrotron emission signals at $\sim$0.5~GeV along the four dataset considered (see Figure~\ref{fig:4SEDs}). 
In turn, the minimum energy of this electron population, $E_{0,\mathrm{tr}}$, is linked to the hardening of the X-ray signal at a few tens of keV. Therefore, if the emission scenario discussed here is correct, variable synchrotron emission originating from turbulent magnetic reconnection can be probed in Mrk~501 by \textit{Swift}/BAT and \textit{Fermi}/LAT, as well as by the forthcoming AMEGO space mission\cite{McEnery_2019}.

The analysis of the cooling timescale
(see Figure~\ref{fig:4cooling}) further indicates that, in the transient-zone,  Fermi acceleration, synchrotron losses, and inverse-Compton (IC) processes occur on timescales shorter than the observed variability 
of approximately one day. 
Notably, the Fermi acceleration timescale in this region is approximately four orders of magnitude shorter than the observed variability.
In the stable zone, these processes mentioned are likewise consistent with the proposed 3-day variability timescale. 
This supports the distinct variability observed in the source during the selected period: the optical-to-soft-X-ray emission components are less variable than the hard-X-ray-to-gamma-ray (TeV)  emission.

In Section 4.4, by exploring the parameter space around the reference model  (Figure \ref{fig:4SEDs}), we could identify how each parameter affects the model’s ability to reproduce the emission observed at MJD 56857.98, particularly the TeV spike. This analysis allowed us to further refine and justify the main parameter choices listed in Tables \ref{table:free_params1} and \ref{table:free_params2}.

For the parameters $\Gamma_\infty$, $f_l$, and $E_{\mathrm{0,b}}$, the values adopted in the reference model (Figure \ref{fig:4SEDs}) appear to be the most appropriate among the ranges explored in the parameter sweep. These values provide a good reproduction of the multi-wavelength SED across the full energy range.

A similar conclusion applies to the location of the transient zone. Although variations in this parameter lead to degeneracies around $\sim 3$ TeV, the first hump of the SED helps constrain its value. In particular, $s_{\mathrm{tr}}=0.20$ pc and $s_{\mathrm{tr}}=0.60$ pc produce clear discrepancies with the SED in the range $10$ keV $< E < 1$ GeV, whereas the value of the reference model, $s_{\mathrm{tr}}=0.35$ pc, provides a better agreement.

 We also find that $s_{\mathrm{tr}}$ and $E_{0,\mathrm{tr}}$ play a key role in shaping the spectral hump near 3 TeV. The steep rise of this SED component is mainly driven by the high minimum electron energy, $E_{0,\mathrm{tr}}$, which suppresses the low-energy SSC emission. Beyond the peak, the sharp spectral change is associated with the onset of the Klein--Nishina regime. This regime strongly affects the cooling of the electrons responsible for the TeV gamma-ray emission, as shown in Figure~\ref{fig:4cooling}, leading to a steepened spectrum.
 At higher energies, the spectrum is further steepened by the cooling-limited cutoff at $E_{\max,\mathrm{tr}}$, together with internal $\gamma\gamma$ attenuation.

In addition, we performed optimized fits for the four datasets using a $\chi^2$-minimization approach (Section 4.5). The resulting fits are broadly consistent with those of the reference model (Figure \ref{fig:4SEDs}), with the main differences involving $E_{0,\mathrm{b}}$, $E_{0,\mathrm{tr}}$, and $f_{\mathrm{v,tr}}$.

$E_{0,\mathrm{b}}$ is a common parameter across the selected observational datasets and has a strong influence on the overall spectral shape, since the stable emitting zone spans a wide energy range. Therefore, a larger variation in this parameter is expected when fitting all datasets simultaneously.
The optimized values of $E_{0,\mathrm{tr}}$ and $f_{\mathrm{v,tr}}$ also show larger deviations, especially for MJD 56857.98 and MJD 56859.97, which are the datasets assigned with higher fitting weights (see ~\ref{appendix_optim}). As a result, the optimization gives priority to these datasets, leading to less satisfactory fits for MJD 56856.91 and MJD 56858.98 compared with the assumptions adopted in the reference model (Figure \ref{fig:4SEDs}). Nevertheless, this comparison shows that the parameters chosen a priori to construct the reference-model SED remain close to the best-fit range for the full dataset.

We note that the SED modeling
requires relatively high low-energy cutoffs for the electron
distributions, with $\gamma_{\min}\gtrsim 10^4$ in the base component and
$\gamma_{\min}\gtrsim 10^5$ in the transient component. We interpret such
$\gamma_{\min}$ as an effective low-energy cutoff of the non-thermal particles
that are efficiently reaccelerated, in the emitting region.

In this picture, the electrons responsible for the emission days considered are not
accelerated directly from the thermal pool of the background plasma.
Rather, they are drawn from a pre-existing relativistic population produced 
by previous dissipation episodes in the jet, also driven by reconnection.
A compact reconnection region can then reaccelerate these seed
particles, shifting the effective low-energy cutoff of the radiating population
to high values. This interpretation is natural in a turbulent, magnetically
dominated jet, where current-driven instabilities  generate turbulence that can 
produce
multiple reconnection sites and allow particles to experience repeated episodes
of acceleration \citep[e.g.][]{Kadowaki2021,medina2021particle, medina2023particle,de2024particle}.

A simple order-of-magnitude estimate shows that such pre-acceleration is feasible.
For instance, the pre-acceleration time is much shorter than on the larger scales considered in the model, because this time scales with the thickness of the reconnection layer. During pre-acceleration, this thickness is at most of the order of the Larmor radius associated with the low-energy cutoff, $r_{L,\rm pre}$. Thus,  
$t_{\rm acc,pre} \sim (r_{L,\rm pre}/\Delta) \times t_{\rm acc}$, where $t_{\rm acc}$ is the acceleration time   (eq.\ref{t_acc_e}) and $\Delta$  the corresponding thickness of the reconnection layer adopted in the reference model for the two regions (eq. \ref{tacc_Xu}). This thickness is of course, much larger than $r_{L,\rm pre}$ and thus, $t_{\rm acc,pre} <<  t_{\rm acc}$, for both regions of the model.

This interpretation is also consistent with the high-energy state of Mrk~501
during the 2014 campaign. The source was already in an active X-ray and
very-high-energy $\gamma$-ray state days before the appearance of the TeV spike \cite{MAGIC_2020}.
This suggests that the jet was undergoing
particle acceleration before the transient component modelled here. The large value of $\gamma_{\min}$ in the transient zone may
therefore reflect the reprocessing of an already energetic electron population
by a compact reconnection event, rather than the temperature of the bulk plasma.

Overall, compared to previous works, the main advantage of the present scenario is not simply that it reproduces the peculiar TeV spike, but that it does so within a physically motivated jet structure. Unlike previous interpretations that were designed mainly for the spike day alone, our model is applied consistently to four consecutive SEDs, including the days before and after the event. In contrast to ad hoc two-zone SSC approaches, the two emitting regions are not assigned arbitrarily, but are tied to a striped-jet turbulence-driven reconnection framework in which their location, magnetization, bulk Lorentz factor, and size follow from the jet evolution. Different to the EED pile-up \cite{MAGIC_2020} and leptohadronic alternatives \cite{petropoulou2024tev}, our model explicitly checks that acceleration and cooling times remain consistent with the observed variability. Therefore, what makes the present model stand out is that it provides a unified physical interpretation for the simultaneous X-ray hardening and TeV spike, with fewer arbitrary assumptions about the emitting zones and with a direct connection between particle acceleration, cooling, jet dynamics, and day-to-day spectral evolution.

The present results highlight the strong potential of turbulent-reconnection--driven Fermi acceleration to produce VHE flares accompanied by an X-ray spectral hardening in high-peaked BL Lac objects.

\section*{Declaration of generative AI and AI-assisted technologies in the manuscript preparation process}
During the preparation of this work, the author(s) used ChatGPT, DeepSeek, and Grammarly in order to improve spelling and grammar. After using this tool/service, the author(s) reviewed and edited the content as needed and take(s) full responsibility for the content of the publication.

\section*{Acknowledgements}
JGGFP  acknowledges the support from the Brazilian Federal Agency for Support and Evaluation of Graduate Education CAPES (grant 88887.600287/2021-00);
EMdGDP acknowledges support support from the São Paulo state funding agency FAPESP (grants 2013/10559-5 and 2021/02120-0) and from CNPq Productivity Grant (308643/2017-8); JCRR acknowledges support from Rio de Janeiro State Funding Agency FAPERJ (grant E-26/205.635/2022); UBA acknowledges support from a Rio de Janeiro State Funding Agency FAPERJ (grants E-26/200.532/2023 and 260003/015700/202) and a CNPq Productivity Grant 309053/2022-6. The authors thank the MAGIC collaboration for providing the data from \cite{MAGIC_2020}. 


\appendix

\section{
SED fitting with $\chi^2$ minimization}
\label{appendix_optim}
To obtain the optimal model parameters of the blazar-turbulent-reconnection scenario using the four-day SED data sets of Mrk501 discussed in the paper,
we propose the minimization of a {\it joint} multi-objective function of the form
\begin{equation}
\mathcal{F}({\bf \Theta}) = \sum_i w_i\chi_i^2,
\label{F_joint}
\end{equation}
where we try to minimize the $\chi^2$ contributions of each 
dataset in order to obtain the overall best fit for the four days.

In equation \ref{F_joint}, $w_i$ is the weight assigned to the $i-$th dataset. These weights are introduced to prioritize the fit to specific individual datasets while reducing the computational cost of the search. Here we set
$\bar{w} = [1,4,1,4]$, prioritizing the fit to dataset of MJD 56857.98 
which shows the clearest TeV feature, and dataset of MJD 56859.97, 
which shows the simultaneous disappearance of both the TeV feature and the hard X-ray enhancement.

In equation (\ref{F_joint}), 
each objective function corresponds to the standard $\chi^2_i$ function for a single data set:
\begin{equation}
\chi^2_i = 
\sum_j^{N_\mathrm{p}}
\left(
\frac{F_j - F_{\mathrm{mod}}({\bf 
 \Pi}_i;E_j)}{\sigma_j}
\right)^{2}.
\label{single_chi2}
\end{equation}
where $F_j$ is the flux value of the $j-$th data point at the emission energy $E_j$, $\sigma_j$ is its associated flux uncertainty, and $F_{\mathrm{mod}, j}$ is the emission flux predicted by the blazar-turbulent-reconnection model described in the article (Figure 2). This model depends on the photon energy $E_j$ and on eleven model parameters, namely:
\begin{equation}
{\bf \Pi}_i  = 
[\Gamma_\infty,\, f_\ell,\, f_{v,b},\, \eta_\mathrm{b},\, E_\mathrm{0,b},\,\alpha_\mathrm{b},\, s_\mathrm{tr},\, \alpha_\mathrm{tr},
        f_{v,\mathrm{tr}}^{(i)},\, \eta_\mathrm{tr}^{(i)},\, E_{0,\mathrm{tr}}^{(i)}].
\end{equation}

Therefore, the multi-objective function in equation \ref{F_joint} depends on a parameter vector ${\bf \Theta}$ composed of ``global'' and ``local'' parameters:
\begin{align}
\boldsymbol{\Theta}= &\bigl[\underbrace{
\Gamma_\infty,\, f_\ell,\, f_{v,b},\, \eta_b,\, E_{0,b},\,
\alpha_b,\, s_\mathrm{tr},\, \alpha_\mathrm{tr}
}_{\text{globals (8)}},\\
& \underbrace{f_{v,\mathrm{tr}}^{(1)},\, \eta_\mathrm{tr}^{(1)},\, E_{0,\mathrm{tr}}^{(1)}
}_{\text{dataset 1}} 
, 
\underbrace{
f_{v,\mathrm{tr}}^{(2)},\, \eta_\mathrm{tr}^{(2)},\, E_{0,\mathrm{tr}}^{(2)}
}_{\text{dataset 2}}
\nonumber
, 
\underbrace{
f_{v,\mathrm{tr}}^{(3)},\, \eta_\mathrm{tr}^{(3)},\, E_{0,\mathrm{tr}}^{(3)}
}_{\text{dataset 3}} 
, 
\underbrace{
f_{v,\mathrm{tr}}^{(4)},\, \eta_\mathrm{tr}^{(4)},\, E_{0,\mathrm{tr}}^{(4)}
}_{\text{dataset 4}} 
\bigr].
\end{align}

The global parameters are free parameters shared by the four datasets, reflecting the quasi-stationary nature of the base emission zone.
The local parameters are free parameters that vary from one dataset to another and describe the behavior of the transient emission component. Hence, the total parameter vector $\boldsymbol{\Theta}$ has dimension
$N_\mathrm{c} = 8 + 4\times3 = 20$.
We minimize the multi-objective function \eqref{F_joint} using the Differential Evolution (DE) algorithm \cite{Storn1996}, with a mutation factor of $F=0.5$, a crossover probability of $C_r=0.7$, and a total population of $N_0 = 15$.

We perform this minimization procedure within the following parameter ranges:
\begin{itemize}
\item 
$\Gamma_\infty \in$ [20, 30],
\item
$f_\ell \in$ [99, 150],
\item 
$f_\mathrm{v,b}\in$  [0.1, 0.9],
\item 
$\eta_\mathrm{b}\in$ [$10^{-3}$, $10^{-2}$],
\item
$E_\mathrm{0,b}\in$ [5.0,  50.0] $\times 10^3m_\mathrm{e}c^2$,
\item 
$\alpha_\mathrm{b}\in[2.5, 3.0]$,
\item 
$s_\mathrm{tr}$[pc] $\in$ [0.05, 0.5],
\item 
$f_\mathrm{v,tr}\in$ [$10^{-3}$, $10^{-2}$],
\item 
$\eta_\mathrm{tr}\in$ [$10^{-4}$, $10^{-2}$],
\item 
$E_\mathrm{0,tr}\in$ [5.0 , 70.0] $\times 10^4 m_\mathrm{e}c^2$,
\item 
$\alpha_\mathrm{tr}\in$ [2.5, 3.0],
\end{itemize}
which enclose the parameter values of the reference model adopted in Section 4.1.

\begin{figure}
   \centering
   \includegraphics[width=\hsize]{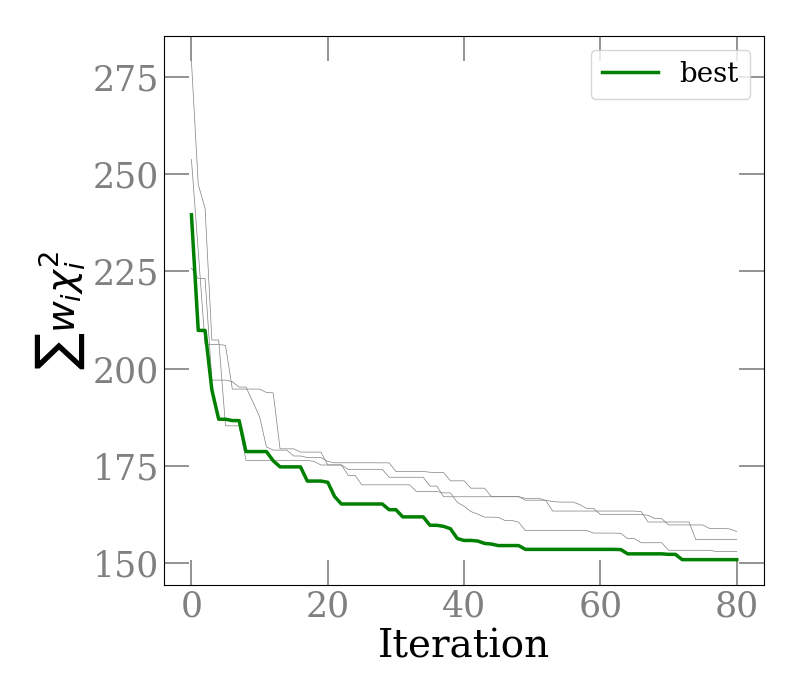}
\caption{
Evolution of the multi-objective function $\mathcal{F}$ given by equation \ref{F_joint}, employed to fit the four SED data sets with the blazar-turbulent-reconnection model discussed in the paper.
The curves show the minimization process performed with the DE algorithm up to 80 iterations.
Each curve represents an independent realization of the minimization process.}
         \label{fig:convergences}
\end{figure}

In Figure~\ref{fig:convergences}, we show the convergence curves of four independent minimization realizations, displaying the evolution of the function $\mathcal{F}$ over 80 iterations.  
The components of the vector ${\bf\Theta}$, i.e., the model parameters, obtained in the last iteration are reported in Tables~\ref{tab:globalopt} and \ref{tab:localopt}.

\begin{table*}
\large
\centering
\caption{Global fitting parameters resulting from the $\chi^2$ minimization test described in Section 4.5. 
These are free parameters common to
the four datasets. Each run is a realization of the minimization procedure that results after 80 iterations. 
}
\begin{tabular}{l c c c c}
\hline
Parameter & Run 1 & Run 2 & Run 3 & Run 4\\
\hline
$\mathcal{F}(\boldsymbol{\Theta})$ & 150.9200 & 153.0100 & 156.0600 & 158.1500 \\
$\Gamma_\infty$ & 25.4690 & 25.4260 & 25.5050 & 25.5280 \\
$f_l$ & 99.8090 & 99.2200 & 99.7100 & 99.7850 \\
$f_{v,b}$ & 0.1772 & 0.1791 & 0.1801 & 0.1852 \\
$\eta_b$ $[10^{-3}]$ & 2.0423 & 2.0443 & 2.0390 & 2.0592 \\
$E_{0,b}$ $[10^{4} m_e c^2]$ & 2.0914 & 2.0101 & 2.0861 & 1.9813 \\
$\alpha_b$ & 2.6200 & 2.6203 & 2.6200 & 2.6204 \\
$s_\mathrm{tr}$ & 0.3886 & 0.3800 & 0.3987 & 0.3895 \\
$\alpha_\mathrm{tr}$ & 2.6200 & 2.6256 & 2.6294 & 2.6200 \\
\hline
\end{tabular}
\label{tab:globalopt}
\end{table*}

\begin{table*}
\centering
\caption{
Local fitting parameters resulting from the same $\chi^2$ minimization procedure leading to the parameter values in Table~\ref{tab:globalopt}.
}
\begin{tabular}{l c c c c}
\hline
Parameter & MJD 56 & MJD 57 & MJD 58 & MJD 59 \\
\hline
\multicolumn{5}{c}{\textbf{Run 1}} \\
\hline
$f_{v,\mathrm{tr}}\,[10^{-3}]$ & 1.8320  & 2.1993 & 1.9143 & 1.0000 \\
$\eta_\mathrm{tr}\,[10^{-3}]$ & 1.9065 & 2.5577 & 1.8650  &  0.1066 \\
$E_{0,\mathrm{tr}}$ [$10^5 m_e c^2$]& 4.9011 & 4.5956 & 3.2427 & 6.6526  \\

\hline
\multicolumn{5}{c}{\textbf{Run 2}} \\
\hline
$f_{v,\mathrm{tr}}\,[10^{-3}]$ &  1.7768 & 2.0680 & 1.9763  &  1.0000  \\
$\eta_\mathrm{tr}\,[10^{-3}]$ & 2.1261  & 2.6841 & 2.0914 & 0.1082 \\
$E_{0,\mathrm{tr}}$ [$10^5 m_e c^2$]& 5.2778 & 5.0488 & 4.3974 & 6.5877
\\

\hline
\multicolumn{5}{c}{\textbf{Run 3}} \\
\hline
$f_{v,\mathrm{tr}}\,[10^{-3}]$ & 1.8534 & 2.1045 & 1.9272 & 1.0000\\
$\eta_\mathrm{tr}\,[10^{-3}]$ & 2.0570 & 2.8863 & 1.6611& 0.1122\\
$E_{0,\mathrm{tr}}$ [$10^5 m_e c^2$]& 4.2744 & 5.1156 & 3.0002 & 4.8217 \\

\hline
\multicolumn{5}{c}{\textbf{Run 4}} \\
\hline
$f_{v,\mathrm{tr}}\,[10^{-3}]$ &1.8220  & 2.2523 & 2.0936 & 1.0000 \\
$\eta_\mathrm{tr}\,[10^{-3}]$ & 2.2280  & 2.7344 &2.3962 & 0.1039 \\
$E_{0,\mathrm{tr}}$ [$10^5 m_e c^2$] & 4.8271 & 4.7106 & 4.8164 & 5.2092 \\
\hline
\end{tabular}
\label{tab:localopt}
\end{table*}


\bibliographystyle{elsarticle-num} 

\bibliography{refs_edited}






\end{document}